\documentclass[epj,nopacs]{svjour}

\usepackage{graphicx}
\usepackage{bm}
\usepackage{hyperref}
\usepackage{xspace}
\usepackage{lmodern}
\usepackage{microtype}
\usepackage[T1]{fontenc}
\usepackage{amsmath}
\rmfamily
\DeclareFontShape{T1}{lmr}{b}{sc}{<->ssub*cmr/bx/sc}{} \DeclareFontShape{T1}{lmr}{bx}{sc}{<->ssub*cmr/bx/sc}{}

\usepackage{siunitx}
\usepackage{subfig}
\usepackage{booktabs}

\hypersetup{pdftitle={Exploring phase space with Nested Sampling}}

\hypersetup{ colorlinks=true, allcolors=[rgb]{0.26,0.41,0.88}, }

\newcommand{\pvalue}{\text{\textit{p-}value}\xspace}
\newcommand{\pvalues}{\text{\pvalue{}s}\xspace}

\newcommand{\Z}{\ensuremath{\mathcal{Z}}\xspace}

\newcommand{\like}{\ensuremath{\mathcal{L}}\xspace}

\newcommand{\prior}{\ensuremath{\Pi}\xspace}

\newcommand{\Pg}[2]{P\mathopen{}\left(#1\,\rvert\, #2\right)\mathclose{}}

\newcommand{\ofOrder}[1]{\ensuremath{\mathcal{O}\left(#1\right)}\xspace}

\newcommand{\MN}{\textsc{MultiNest}\xspace} 
\newcommand{\PC}{\textsc{PolyChord}\xspace}
\newcommand{\Rivet}{\textsc{Rivet}\xspace}

\newcommand{\term}{\ensuremath{T_C}\xspace}

\newcommand{\likemin}{\ensuremath{\like_{\text{min}}}\xspace}
\newcommand{\nlike}{\ensuremath{N_\like}\xspace}

\newcommand{\nw}{\ensuremath{N_W}\xspace}
\newcommand{\nlive}{\ensuremath{n_\text{live}}\xspace}
\newcommand{\ndim}{\ensuremath{n_\text{dim}}\xspace}
\newcommand{\nrep}{\ensuremath{n_\text{rep}}\xspace}

\newcommand{\nprior}{\ensuremath{n_\text{prior}}\xspace}

\newcommand{\nequals}{\ensuremath{N_\mathrm{equal}}\xspace}
\newcommand{\eff}{\ensuremath{\epsilon}\xspace}
\newcommand{\effss}{\ensuremath{\epsilon_{\text{ss}}}\xspace}
\newcommand{\effuw}{\ensuremath{\epsilon_{\text{uw}}}\xspace}

\newcommand{\pT}{\ensuremath{p_\mathrm{T}}\xspace}
\newcommand{\deltaw}{\ensuremath{\Delta_{w}}\xspace}

\newcommand{\deltatot}{\ensuremath{\Delta\sigma_\mathrm{tot}}\xspace}
\newcommand{\deltamc}{\ensuremath{\Delta_\mathrm{MC}}\xspace}

\newcommand{\Sherpa}{S\protect\scalebox{0.8}{HERPA}\xspace}
\newcommand{\Amegic}{A\protect\scalebox{0.8}{MEGIC}\xspace}
\newcommand{\HAAG}{H\protect\scalebox{0.8}{AAG}\xspace}
\newcommand{\RAMBO}{R\protect\scalebox{0.8}{AMBO}\xspace}
\newcommand{\Vegas}{V\protect\scalebox{0.8}{EGAS}\xspace}
\newcommand{\anesthetic}{\texttt{anesthetic}\xspace}
\newcommand{\nestcheck}{\texttt{nestcheck}\xspace}
\newcommand{\ME}{\ensuremath{\mathcal{M}}\xspace}
\newcommand{\xs}{\ensuremath{\sigma}\xspace}

\def\makeheadbox{}

\begin{document}

\title{Exploring phase space with Nested Sampling}



\author{ David Yallup\inst{1} \and Timo Jan\ss{}en\inst{2} \and Steffen Schumann\inst{2} \and Will Handley\inst{1}}

\institute{Cavendish Laboratory \& Kavli Institute for Cosmology, University of Cambridge, JJ Thomson Avenue, Cambridge, CB3~0HE, United Kingdom\mail{\href{mailto:dy297@cam.ac.uk}{dy297@cam.ac.uk}} \and Institut f\"{u}r Theoretische Physik, Georg-August-Universit\"{a}t {G\"{o}ttingen}, Friedrich-Hund-Platz~1, 37077~{G\"{o}ttingen}, Germany}

\abstract{
We present the first application of a Nested Sampling algorithm to explore the high-dimensional phase space of particle
collision events. We describe the adaptation of the algorithm, designed to perform Bayesian inference computations, to
the integration of partonic scattering cross sections and the generation of individual events distributed according to
the corresponding squared matrix element. As a first concrete example we consider gluon scattering processes into 3-, 4-
and 5-gluon final states and compare the performance with established sampling techniques. Starting from a flat prior
distribution Nested Sampling outperforms the {\sc{Vegas}} algorithm and achieves results comparable to a dedicated
multi-channel importance sampler. We outline possible approaches to combine Nested Sampling with non-flat prior
distributions to further reduce the variance of integral estimates and to increase unweighting efficiencies. 
}

\maketitle

\section{Introduction}\label{sec:intro}

Realistic simulations of scattering events at particle collider experiments play an indispensable role in the analysis
and interpretation of actual measurement data for example at the Large Hadron Collider
(LHC)~\cite{Buckley:2011ms,Campbell:2022qmc}. A central component of such event simulations is the generation of hard
scattering configurations according to a density given by the squared transition matrix element of the concrete process
under consideration. This is needed both for the evaluation of corresponding cross sections, as well as the explicit
generation of individual events that potentially get further processed, \emph{e.g.}\ by attaching parton showers,
invoking phenomenological models to account for the parton-to-hadron transition, and eventually, a detector simulation.
To adequately address the physics needs of the LHC experiments requires the evaluation of a wide range of
high-multiplicity hard processes that feature a highly non-trivial multimodal target density that is rather costly to
evaluate. The structure of the target is thereby affected by the appearance of intermediate resonances, quantum
interferences, the emission of soft and/or collinear massless gauge bosons, or non-trivial phase space constraints, due
to kinematic cuts on the final state particles. Dimensionality and complexity of the phase space sampling problem make
the usage of numerical methods, and in particular Monte Carlo techniques, for its solution indispensable.

The most widely used approach relies on adaptive multi-channel importance sampling, see for
example~\cite{Kleiss:1994qy,Papadopoulos:2000tt,Krauss:2001iv,Maltoni:2002qb,Gleisberg:2008fv}. However, to achieve good
performance detailed knowledge of the target distribution, \emph{i.e.}\ the squared matrix element, is needed. To this
end information about the topology of scattering amplitudes contributing to the considered process is employed in the
construction of individual channels. Alternatively, and also used in combination with importance sampling phase space
maps, variants of the self-adaptive \Vegas algorithm~\cite{Lepage:1977sw} are routinely
applied~\cite{Ohl:1998jn,Jadach:1999sf,Hahn:2004fe,vanHameren:2007pt}.

An alternative approach for sampling according to a desired probability density is offered by Markov Chain Monte Carlo
(MCMC) algorithms. However, in the context of phase space sampling in high energy physics these techniques attracted
rather limited attention, see in particular~\cite{Kharraziha:1999iw,Weinzierl:2001ny}. More recently a mixed kernel
method combining multi-channel sampling and MCMC, dubbed $(\text{MC})^3$, has been presented~\cite{Kroeninger:2014bwa}.
A typical feature of such MCMC based algorithms is the potential autocorrelation of events that can affect their direct
applicability in typical use case scenarios of event generators. 

To meet the computing challenges posed by the upcoming and future LHC collider runs and the corresponding event
simulation campaigns, improvements of the existing phase space sampling and event unweighting techniques will be
crucial~\cite{HSFPhysicsEventGeneratorWG:2020gxw,HSFPhysicsEventGeneratorWG:2021xti}. This has sparked renewed interest
in the subject, largely driven by applications of machine learning techniques, see for
instance~\cite{Bendavid:2017zhk,Klimek:2018mza,Otten:2019hhl,DiSipio:2019imz,Butter:2019cae,Alanazi:2020klf,Alanazi:2020jod,Diefenbacher:2020rna,Butter:2020qhk,Chen:2020nfb,Matchev:2020tbw,Bothmann:2020ywa,Gao:2020vdv,Gao:2020zvv,Stienen:2020gns,Danziger:2021eeg,Backes:2020vka,Bellagente:2021yyh,Butter:2021csz}.

In this article we explore an alternative direction. We here study the application of Nested
Sampling~\cite{Skilling:2006gxv} as implemented in \PC~\cite{Handley:2015vkr} to phase space integration and event
generation for high energy particle collisions. We here assume no prior knowledge about the target and investigate the
ability of the algorithm to adapt to the problem. Nested Sampling has originally been proposed to perform Bayesian
inference computations for high dimensional parameter spaces, providing also the evidence integral, \emph{i.e.}\ the
integral of the likelihood over the prior density. This makes it ideally suited for our purpose. In Sec.~\ref{sec:ns} we
will introduce Nested Sampling as a method to perform cross section integrals and event generation, including a reliable
uncertainty estimation. In Sec.~\ref{sec:gluon} we will apply the method to gluon scattering to $3-$, $4-$, and
$5-$gluon final states as a benchmark for jet production at hadron colliders, thereby comparing results for total cross sections and differential distributions with
established standard techniques. Evaluation of the important features of the algorithm when applied in the particle
physics context is also discussed in this section. In Sec.~\ref{sec:directions} we illustrate several avenues for future
research, extending the work presented here. Finally, we present our conclusions in Sec.~\ref{sec:conc}.

\section{Nested Sampling for event generation}\label{sec:ns}

The central task when exploring the phase space of scattering processes in particle physics is to compute the cross
section integral, \xs. This requires the evaluation of the transition squared matrix element, $|\ME|^2$, integrated over
the phase space volume, $\Omega$, where $\Omega$ is composed of all possible kinematic configurations, $\Phi$, of the
external particles. Up to some constant phase space factors this amounts to performing the integral,
\begin{equation}\label{eq:ps}
    \xs = \int\limits_\Omega d\Phi |\ME|^2 (\Phi)\,.
\end{equation}
In practice rather than sampling the physical phase space variables, \emph{i.e.}\ the particles' four-momenta, it is
typical to integrate over configurations, $\theta\in[0,1]^D$, from the $D$-dimensional unit hypercube. Some mapping,
$\prior:[0,1]^D\to\Omega$, is then employed to translate the sampled variables to the physical momenta. The mapping is
defined as, $\Phi = \prior(\theta)$, and the integral in Eq.~\eqref{eq:ps} is written,
\begin{equation}\label{eq:ps_samp}
    \sigma = \int\limits_{[0,1]^D}  d\theta |\ME|^2 (\prior (\theta)) \mathcal{J}(\theta) =  \int\limits_{[0,1]^D}  d\theta \mathcal{L}(\theta)\,.
\end{equation} 
A Jacobian associated with the change of coordinates between $\theta$ and $\Phi$ has been introduced, $\mathcal{J}$, and
then absorbed into the definition of $\mathcal{L}(\theta) = |\ME|^2(\prior(\theta)) \mathcal{J}(\theta)$. With no
general analytic solution to the sorts of scatterings considered at the high energy frontier, this integral must be
estimated with numerical techniques. Numerical integration involves sampling from the $|\ME|^2$ distribution in a manner
that gives a convergent estimate of the true integral when the samples are summed. As a byproduct this set of samples
can be used to estimate integrals of arbitrary sub-selections of the integrated phase space volume, decomposing the
total cross section into differential cross section elements, $d\sigma$. Additionally these samples can be unweighted
and used as pseudo-data to emulate the experimental observations of the collisions. The current state of the art
techniques for performing these tasks were briefly reviewed in Section~\ref{sec:intro}.

Importance Sampling (IS) is a Monte Carlo technique used extensively in particle physics when one needs to draw samples
from a distribution with an unknown \emph{target} probability density function, $P(\Phi)$. Importance Sampling
approaches this problem by instead drawing from a known \emph{sampling} distribution, $Q(\Phi)$ (A number of standard
texts for inference give more thorough exposition of the general sampling theory used in this paper, see \emph{e.g.}\
\cite{mackay}). Samples drawn from $Q$ are assigned a weight, $w=P(\Phi)/Q(\Phi)$, adjusting the importance of each
sampled point. The performance of IS rests heavily on how well the sampling distribution can be chosen to match the
target, and adaptive schemes like \Vegas are employed to refine initial proposals. It is well established that as the
dimensionality and complexity of the target increase, the task of constructing a viable sampling distribution becomes
increasingly challenging.

Markov Chain based approaches fundamentally differ in that they employ a local sampling distribution and define an
acceptance probability with which to accept new samples. Markov Chain Monte Carlo (MCMC) algorithms are widely used in
Bayesian inference. Numerical Bayesian methods have to be able to iteratively refine the prior distribution to the
posterior, even in cases where the two distributions are largely disparate, making stochastic MCMC refinement an
indispensable tool in many cases. This is an important conceptual point; in the particle physics problems presented in
this work we are sampling from exact theoretically derived distributions. The lack of noise and a priori well known
structure make methods with deterministic proposal distributions such as IS more initially appealing, however at some
point increasing the complexity and dimensionality of the problem forces one to use stochastic methods. Lattice QCD
calculations are a prominent example set of adjacent problems sampling from theoretical distributions that make
extensive use of MCMC approaches~\cite{ParticleDataGroup:2020ssz}. MCMC algorithms introduce an orthogonal set of
challenges to IS; a local proposal is inherently simpler to construct, however issues with exploration of multimodal
target distributions and autocorrelation of samples become new challenges to address.

Nested Sampling (NS) is a well established algorithm for numerical evaluation of high dimensional
integrals~\cite{Skilling:2006gxv}. NS differs from typical MCMC samplers as it is primarily an integration algorithm,
hence by definition has to overcome a lot of the difficulties MCMC samplers face in multimodal problems. A recent
community review of its various applications in the physical sciences, and various implementations of the algorithm has
been presented in~\cite{Ashton:2022grj}. 

\begin{figure}
  \begin{center}
    \includegraphics[width=.7\columnwidth,page=1]{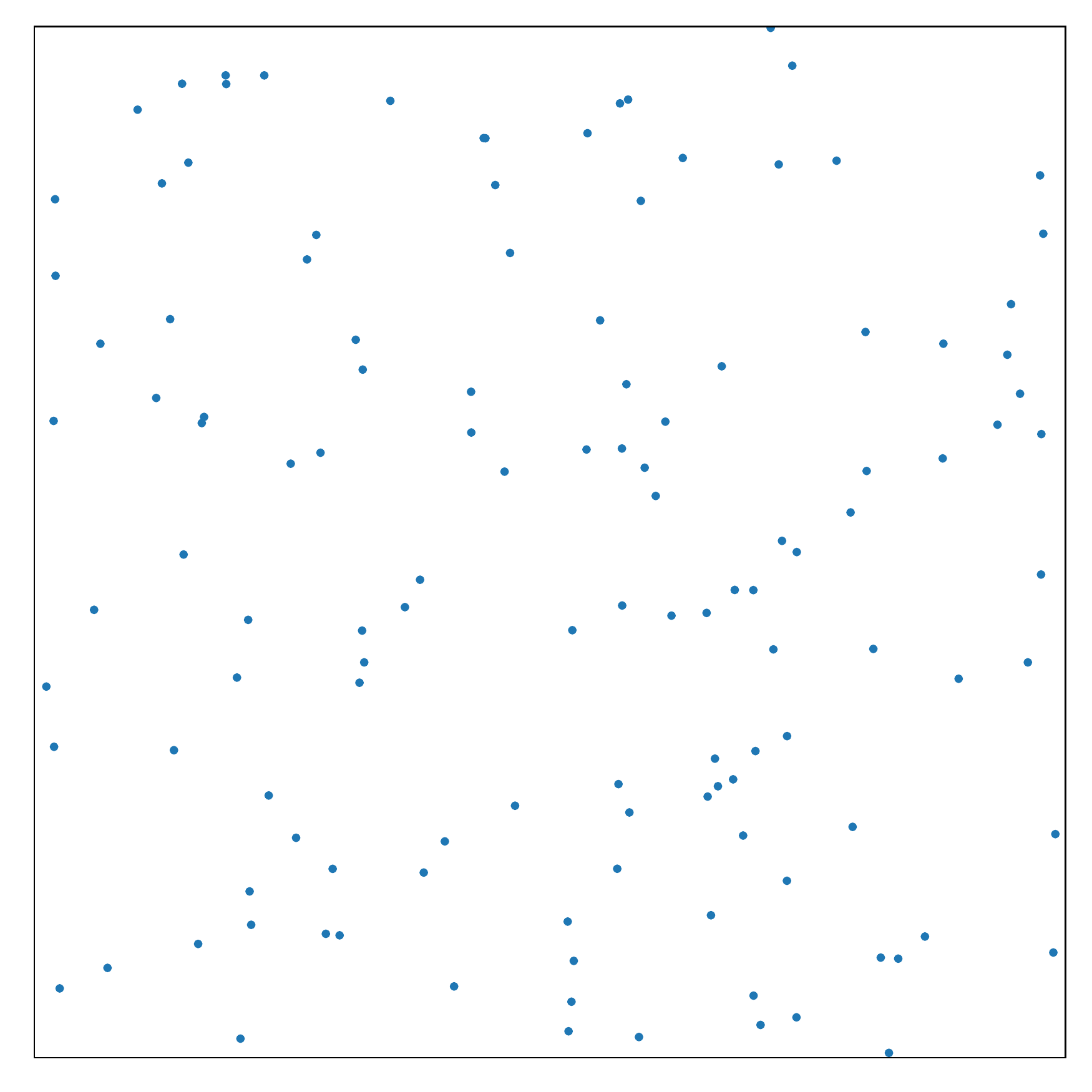}
    \includegraphics[width=.7\columnwidth,page=2]{figures/himmelblau}
    \includegraphics[width=.7\columnwidth,page=6]{figures/himmelblau}
\caption{Schematic of live point evolution (blue dots) in Nested Sampling, over a two-dimensional function whose logarithm is the negative Himmelblau function
(contours). Points are initially drawn from the unit hypercube (top panel). The points on the lowest contours are
successively deleted, causing the live points to contract around the peak(s) of the function. After sufficient
compression is achieved, the dead points (orange) may be weighted to compute the volume under the surface and samples
from probability distributions derived from the function.\label{fig:nsillustration}}
\end{center}
\end{figure}

At its core NS operates by maintaining a number, \nlive, of \textit{live point} samples. This ensemble of live points is
initially uniformly sampled from $\theta\in[0,1]^D$ -- distributed in the physical volume $\Omega$ according to the
shape of the mapping \prior. These live points are sorted in order of $\like(\theta)$ evaluated at the phase space
point, and the point with the lowest \like, \likemin, in the population is identified. A replacement for this point is
found by sampling uniformly under a hard constraint requiring, $\like>\likemin$. The volume enclosed by this next
iteration of live points has contracted and the procedure of identifying the lowest \like point and replacing it is
repeated. An illustration of three different stages of this iterative compression on an example two-dimensional function
are shown in Figure~\ref{fig:nsillustration}. The example function used in this case has four identical local maxima to find, practical exploration and discovery of the modes is achieved by having a sufficient (\ofOrder{10}) initial samples in the basis of attraction of each mode. This can either be achieved by brute force sampling a large number of initial samples, or by picking an initial mapping distribution that better reflects the multi-modal structure. By continually uniformly sampling from a steadily compressing volume, NS
can estimate the density of points which is necessary for computing an integral as given by Eq.~\eqref{eq:ps}. Once the
iterative procedure reaches a point where the live point ensemble occupies a predefined small fraction of the initial
volume, \term, the algorithm terminates. The fraction \term can be characterised as the \textit{termination criterion}.
The discarded points throughout the evolution are termed \textit{dead points} which can be joined with the remaining
live points to form a representative sample of the function, that can be used to estimate the integral or to provide a
random sample of events. 

To estimate the integral and generate (weighted) random samples, Nested Sampling achieves this by probabilistically
estimating the volume of the shell between the two outermost points as approximately $\frac{1}{\nlive}$ of the current
live volume. The volume $X_j$ within the contour $\mathcal{L}_j$ -- defined by the point with $\likemin$ -- at iteration
$j$ may therefore be estimated as,
\begin{equation*}
  \begin{split}
    X_j = \int_{\mathcal{L}(\theta)>\mathcal{L}_j}d\theta \qquad\Rightarrow\qquad {}&  X_0=1,  \\
    \quad P(X_{j}|X_{j-1})  = \frac{X_j^{\nlive-1}}{\nlive X_{j-1}^{\nlive}} \qquad\Rightarrow\qquad  {}&\log X_j \approx \frac{-j \pm \sqrt{j}}{\nlive}.
  \end{split}
\end{equation*}
The cross section and probability weights can therefore be estimated as,
\begin{equation}
  \begin{split}
     \sigma = {}& \int d\theta \mathcal{L}(\theta) = \int dX \mathcal{L}(X) \\
    \approx {}&  \sum_j \mathcal{L}_j \Delta X_j, \qquad w_j \approx \frac{\Delta X_j\mathcal{L}_j }{\sigma}.
    \end{split} 
\end{equation}
Importantly, for all of the above the approximation signs indicate errors in the procedure of probabilistic volume
estimation, which are fully quantifiable.

The method to sample new live points under a hard constraint can be realised in multiple ways, and this is one of the
key differences in the various implementations of NS. In this work we employ the \PC implementation of Nested
Sampling~\cite{Handley:2015vkr}, which uses slice sampling~\cite{nealslicesampling} MCMC steps to evolve the live
points. NS can be viewed as being an ensemble of many short Markov Chains.

Much of the development and usage of NS has focused on the problem of calculation of marginal likelihoods (or evidences)
in Bayesian inference, particularly within the field of
Cosmology~\cite{Mukherjee:2005wg,Shaw:2007jj,Feroz:2007kg,Feroz:2008xx,Feroz:2013hea,Handley:2015fda}. We can define the
Bayesian evidence, \Z, analogously to the particle physics cross section, \xs. NS in this context evaluates the
integral,
\begin{equation}\label{eq:evidence}
  \Z = \int d\theta \like (\theta) \pi (\theta)\,,
\end{equation}
where the likelihood function, \like, plays a similar role to $|\ME|^2$. In the Bayesian inference context, the
phase space over which we are integrating, $\theta$, has a measure defined by the prior distribution, $\pi(\theta)$,
which without loss of generality under a suitable coordinate transformation can be taken to be uniform over the unit
hypercube. Making the analogy between the evidence and the cross section explicit will allow us to apply some of the
information theoretic metrics commonly used in Bayesian inference to the particle physics
context~\cite{Handley:2019pqx}, and provide terminology used throughout this work. Among a wide array of sampling
methods for Bayesian inference, NS possesses some unique properties that enable it to successfully compute the high
dimensional integral associated with Eq.~\eqref{eq:evidence}. These properties also bear a striking similarity to the
requirements one would like to have to explore particle physics phase spaces. These are briefly qualitatively described
as follows:

\begin{itemize}
\item NS is primarily a \emph{numerical integration method that produces posterior samples as a by product}. In this
respect it is comfortably similar to Importance Sampling as the established tool in particle physics event generation.
It might initially be tempting to approach the particle physics event generation task purely as a posterior sampling
problem. Standard Markov Chain based sampling tools cannot generically give good estimates of the integral, so are not
suited to compute the cross section. Additionally issues with coverage of the full phase space from the resulting event
samples are accounted for by default by obtaining a convergent estimate of the integral over all of the phase space. 
\item NS naturally \emph{handles multimodal problems}~\cite{Feroz:2007kg,Feroz:2008xx}. The iterative compression can be
augmented by inserting steps that cluster the live points periodically throughout the run. Defining subsets of live
points and evolving them separately allows NS to naturally tune itself to the modality of unseen problems.
\item NS requires a construction that can handle \emph{sampling under a hard likelihood constraint} in order to perform
the compression of the volume throughout the run. Hard boundaries in the physics problem, such as un-physical or
deliberately cut phase space regions, manifest themselves in the sampling space as a natural extension of these
constraints.
\item NS is \emph{largely self tuning}. Usage in Bayesian inference has found that NS can be applied to a broad range of
problems with little optimisation of hyper-parameters
necessary~\cite{AbdusSalam:2020rdj,Martinez:2017lzg,Fowlie:2020gfd}. NS can adapt to different processes in particle
physics \emph{without any prior knowledge of the underlying process needed}.
\end{itemize}

The challenge to present NS in this new context is to find an even comparison of sampling performance between NS and IS.
It is typical in phase space sampling to compare the difference between the target and the sampling distribution as
reducing the variation between these two distributions gives a clear metric of performance for IS. For NS there is no
such global sampling distribution; the closest analogue being the prior which is then iteratively refined with local
proposals to an estimate of the target. In Section~\ref{sec:toy_example} we attempt to compare the sampling distribution
between NS and IS using a toy problem, however in the full physical gluon scattering example presented in
Section~\ref{sec:gluon} we instead focus directly on the properties of the estimated target distribution as this is the
most direct equitable point of comparison.

\subsection{Illustrative example}\label{sec:toy_example}

To demonstrate the capabilities of NS we apply the algorithm to an illustrative sampling problem in two dimensions.
Further examples validating \PC on a number of challenging sampling toy problems are included in the original
paper~\cite{Handley:2015vkr}, here we present a modified version of the \emph{Gaussian Shells} scenario. An important
distinction of the phase space use case not present in typical examples is the emphasis on calculating finely binned
\emph{differential} histograms of the total integral. As a comparison to NS, we sample the same problem with a method
that is well-known in high energy physics -- adaptive Importance Sampling (IS), realised using the \Vegas algorithm.

For our toy example we introduce a ``stop sign'' target density, whose unnormalised distribution is defined by
\begin{equation}
  \begin{split}
    f(x, y) ={}& \frac{1}{2\pi^2} \frac{\Delta r}{\left(\sqrt{(x-x_0)^2+(y-y_0)^2}-r_0\right)^2+(\Delta r)^2} \\
    {}& \cdot \frac{1}{\sqrt{(x-x_0)^2+(y-y_0)^2}}\\
    {}& + \frac{1}{2\pi r_0} \frac{\Delta r}{((y-y_0)-(x-x_0))^2+(\Delta r)^2} \\
    {}& \cdot \Theta\left(r_0 - \sqrt{(x-x_0)^2+(y-y_0)^2}\right)\,,
  \end{split}
\end{equation}
where $\Theta(x)$ is the Heaviside function. It is the sum of a ring and a line segment, both with a (truncated) Cauchy
profile. The ring is centred at $(x_0, y_0) = (0.5, 0.5)$ and has a radius of $r_0 = 0.4$. The line segment is located
in the inner part of the ring and runs through the entire diameter. We set the width of the Cauchy profile to $\Delta r
= 0.002$. This distribution can be seen as an example of a target where it makes sense to tackle the sampling problem
with a multi-channel distribution. One channel could be chosen to sample the ring in polar coordinates and one to sample
the line segment in Cartesian coordinates. However, here we deliberately use \Vegas as a single channel in order to
highlight the limitations of the algorithm. From the perspective of a single channel, there is no coordinate system to
factorise the target distribution. That poses a serious problem for \Vegas, as it uses a factorised sampling
distribution where the variables are sampled individually. Both algorithms are given zero prior knowledge of the target,
thus starting with a uniform prior distribution.

\begin{figure}
  \begin{center}
    \subfloat[target]{%
      \hspace{.3cm}
      \includegraphics[width=.3564\textwidth]{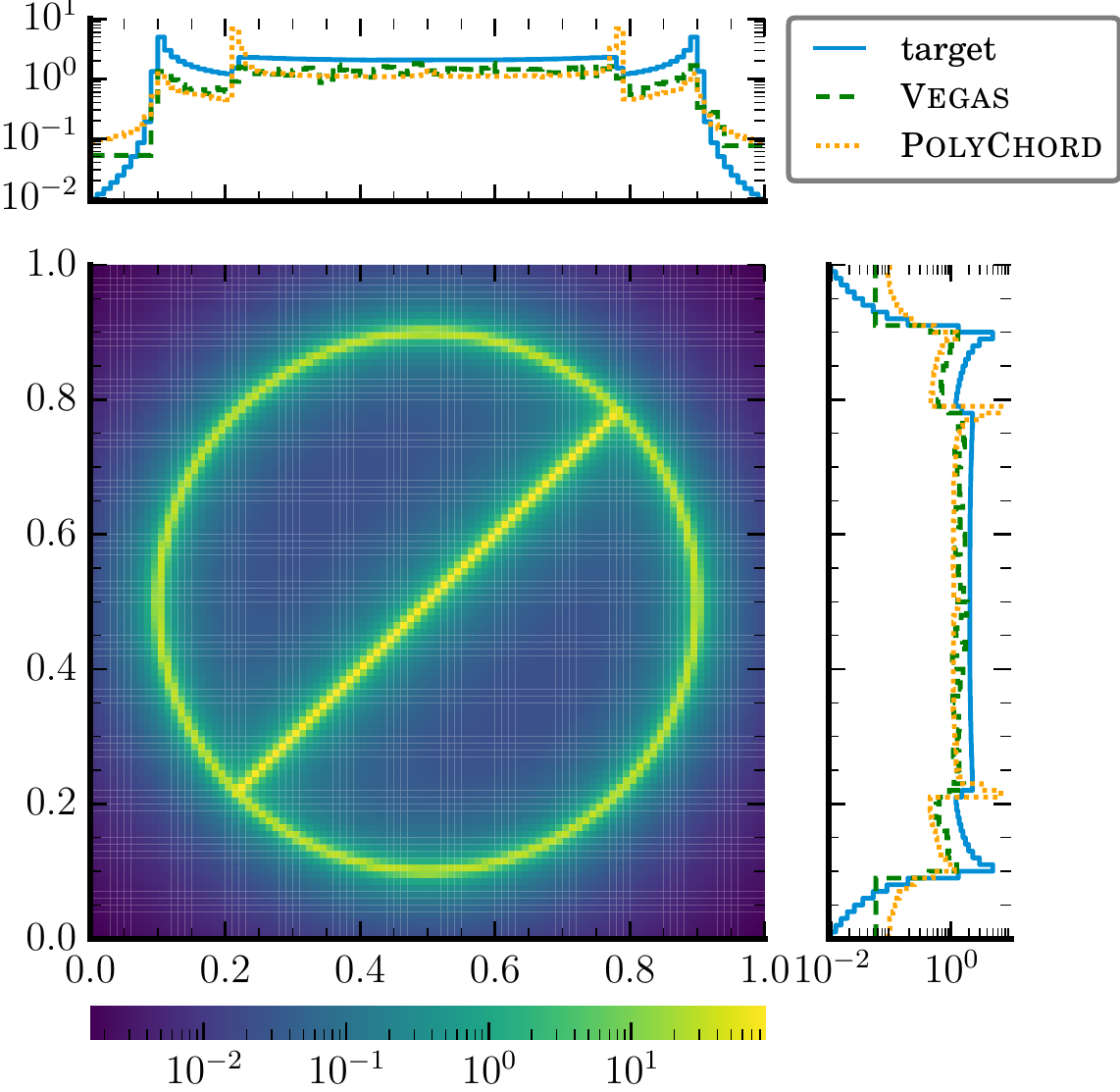}%
      \label{fig:toy_target}%
    }
    \\
    \subfloat[ratio target/\Vegas sampling density]{%
      \includegraphics[width=0.3168\textwidth]{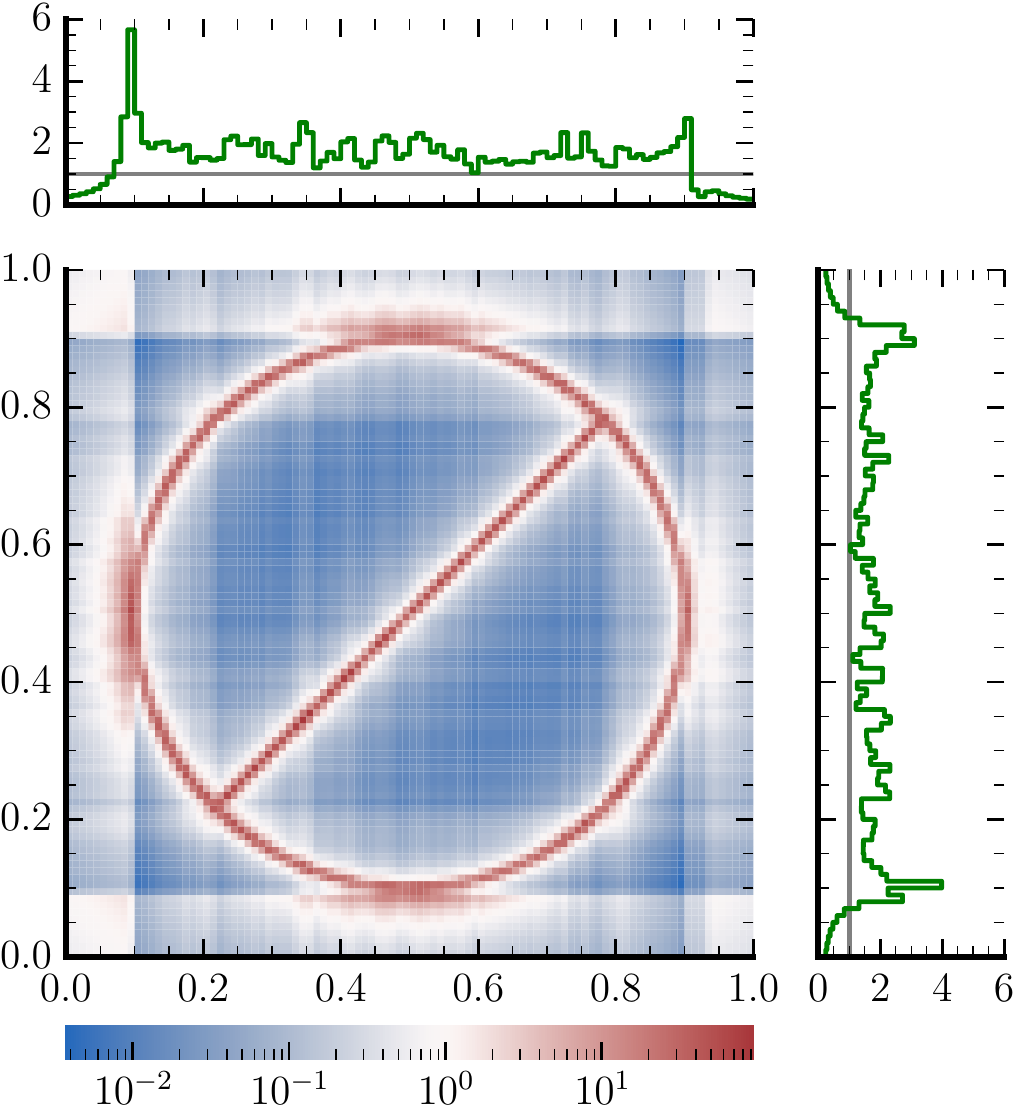}%
      \label{fig:toy_vegas}%
    }
    \\    
    \subfloat[ratio target/\PC sampling density]{%
      \includegraphics[width=0.3168\textwidth]{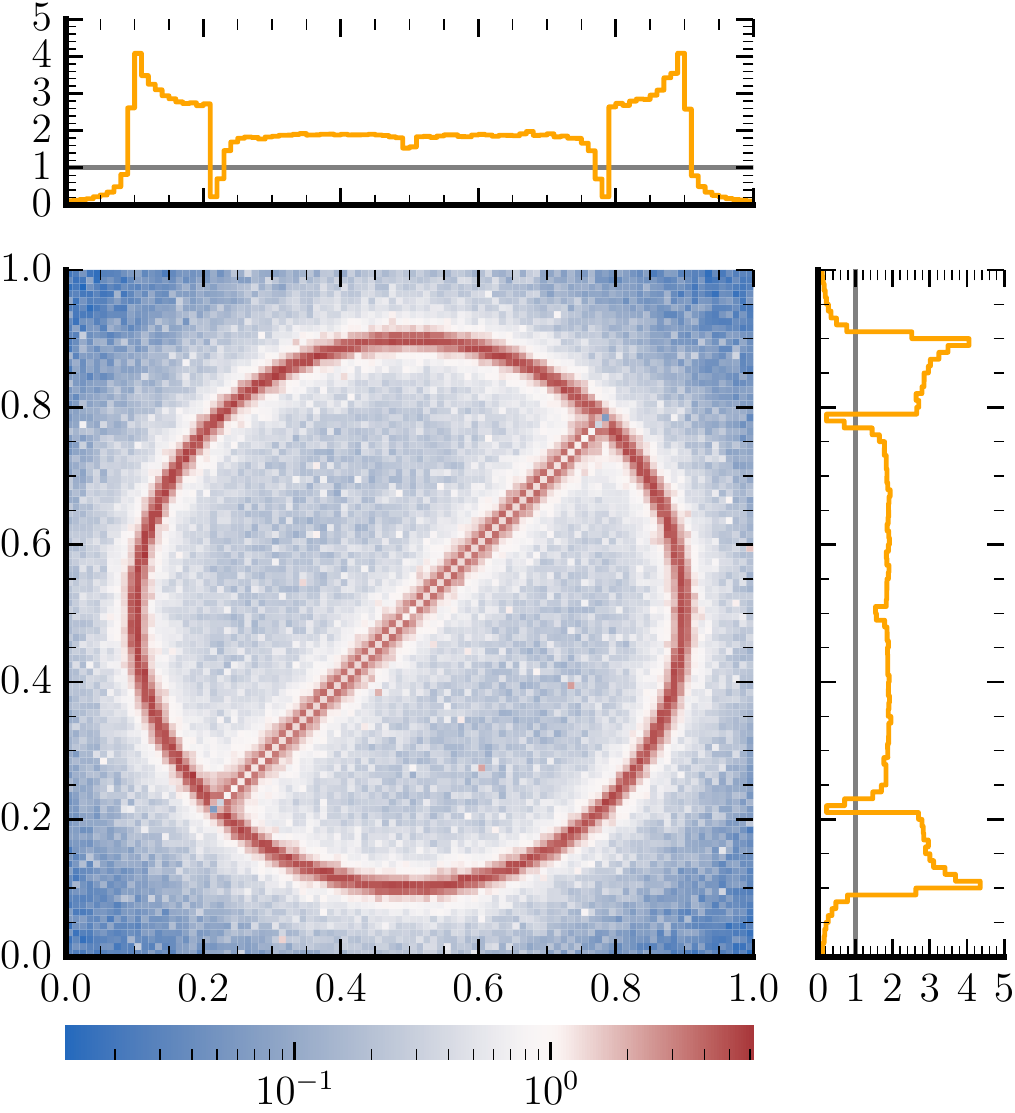}%
      \label{fig:toy_polychord}%
    }
    \caption{A two-dimensional toy example: \protect\subref{fig:toy_target} Histogram of the target function along with
the marginal sampling distributions of \Vegas and \PC. \protect\subref{fig:toy_vegas} Ratio of the target function and
the probability density function of \Vegas. \protect\subref{fig:toy_polychord} Ratio of the target density to the
sampling density of \PC.
    }
    \label{fig:toy}
  \end{center}
\end{figure}

Our \Vegas grid has 200 bins per dimension. We train it over 10 iterations where we draw 30k points from the current
\Vegas mapping and adapt the grid to the data. The distribution defined by the resulting grid is then used for IS
without further adaptation. This corresponds to the typical use in an event generator, where there is first an
integration phase in which, among other things, \Vegas is adapted, followed by a non-adaptive event generation phase. We
note that \Vegas gets an advantage in this example comparison as we do not include the target evaluations from the
training into the counting. However, it should be borne in mind that in a realistic application with a large number of
events to be generated, the costs for training are comparatively low. For NS we use \PC with a number of live points
$n_\text{live} = \num{1000}$ and a chain length $n_\text{repeats} = 4$, more complete detail of \PC settings and their
implication are given in Section~\ref{subsec:hyper}. Fig.~\ref{fig:toy_target} shows the bivariate target distribution
along with the marginal $x$ and $y$ distributions of the target, \Vegas and \PC. For this plot (as well as for
Fig.~\ref{fig:toy_polychord}) we merged 70 independent runs of \PC to get a better visual representation due to the
larger sample size. It can be seen that both algorithms reproduce the marginal distributions reasonably well. There is
some mismatch at the boundaries for \Vegas. This can be explained by the fact that \Vegas, as a variance-reduction
method, focuses on the high-probability regions, where it puts many bins, and uses only few bins for the comparably flat
low-probability regions. As a result, the bins next to the boundaries are very wide and overestimate the tails. \PC also
oversamples the tails, reflecting the fact that in this example the prior is drastically different from the posterior,
meaning the initial phase of prior sampling in \PC is very inefficient. In addition it puts too many points where the
ring and the line segment join, which is where we find the highest values of the target function. This is not a generic
feature of NS at the termination of the algorithm, rather it reflects the nature of having two intersecting sweeping
degenerate modes in the problem, a rather unlikely scenario in any physical integral.

Fig.~\ref{fig:toy_vegas} shows the ratio between the target distribution and the sampling distribution of \Vegas,
representing the IS weights. It can be seen that the marginals of the ratio are relatively flat, with values between
\num{0.1} and \num{5.7}. However, in two dimensions the ratio reaches values up to \num{1e2}. By comparing
Fig.~\ref{fig:toy_target} and Fig.~\ref{fig:toy_vegas}, paying particular attention to the very similar ranges of
function values, it can be deduced that \Vegas almost completely misses to learn the structure of the target. It tries
to represent the peak structure from the ring and the line segment by an enclosing square with nearly uniform
probability distribution.

\begin{figure}
  \begin{center}
  \includegraphics[]{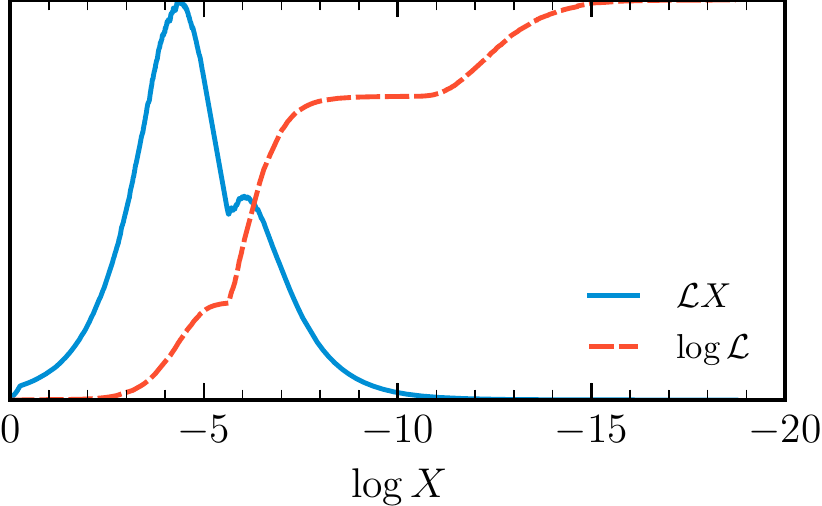}
\caption{Likelihood ($\log\like$) and posterior mass ($\like X$) profiles for a run of \PC on the example target
density. The $x$-axis tracks the prior volume remaining as the run progresses, with $\log X=0$ corresponding to the
start of the run, with the algorithm compressing the volume from left to right, where the run
terminates.}\label{fig:higson}

\end{center}
\end{figure}

The same kind of plot is shown in Fig.~\ref{fig:toy_polychord} for the \PC data. NS does not strictly define a sampling
distribution, however a proxy for this can be visualised by plotting the density of posterior samples. Here the values
of the ratio are much smaller, between \num{1e-2} and \num{7}. \PC produces a flatter ratio function than \Vegas while
not introducing additional artifacts that are not present in the original function. The smallest/largest values of the
ratio are found in the same regions as the smallest/largest values of the target function, implying that \PC tends to
overestimate the tails and to underestimate the peaks. This can be most clearly explained by examining the profile of
where posterior mass is distributed throughout a run, an important diagnostic tool for NS runs~\cite{Higson_2018}. It is
shown in Fig.~\ref{fig:higson}, where the algorithm runs from left to right; starting with the entire prior volume
remaining enclosed by the live points, $\log X=0$, and running to termination, when the live points contain a
vanishingly small remaining prior volume. The posterior mass profile, shown in blue, is the analogue to the sampling
density in \Vegas. To contextualise this against the target function, a profile of the log-likelihood of the lowest live
point in the live point ensemble is similarly shown as a function of the remaining prior volume, $X$. Nested Sampling
can be motivated as a \emph{likelihood scanner}, sampling from monotonically increasing likelihood shells. These two
profiles indicate some features of this problem, firstly a \emph{phase transition} is visible in the posterior mass
profile. This occurs when the degenerate peak of the ring structure is reached, the likelihood profile reaches a plateau
where the iterations kill off the degenerate points at the peak of the ring, before proceeding to scan up the remaining
line segment feature. An effective second plateau is found when the peak of the line segment is reached, with a final
small detail being the superposition of the ring likelihood on the line segment. Once the live points are all occupying
the extrema of the line segment, there is a sufficiently small prior volume remaining that the algorithm terminates. The
majority of the posterior mass, and hence sampling density is distributed around the points where the two peaks are
ascended. This reflects the stark contrast between the prior initial sampling density and the target, the samples are
naturally distributed where the most information is needed to effectively compress the prior to the posterior. 

We compare the efficiencies of the two algorithms for the generation of equal-weight events in Tab.~\ref{tab:toy_eff}.
It shows that \PC achieves an overall efficiency of $\eff = \num{0.0113 +- 0.009}$ which is almost three times as high
as the efficiency of \Vegas. While for \Vegas the overall efficiency $\eff$ is identical to the unweighting efficiency
$\effuw$, determined by the ratio of the average event weight over the maximal weight in the sample, for \PC we also
have to take the slice sampling efficiency $\effss$ into account, which results from the thinning of the Markov Chain in
the slice sampling step. Here, the total efficiency $\eff = \effss \effuw$ is dominated by the slice sampling
efficiency. We point out that it is in the nature of the NS algorithm that the sample size is not deterministic.
However, the variance is not very large and it is easily possible to merge several NS runs to obtain a larger sample.
\begin{table*}
  \centering
\caption{Comparison of \Vegas and NS for the toy example in terms of size of event samples produced. \nlike gives the
number of target evaluations, \nw the number of weighted events and \nequals the derived number of equal weight events.
A MC slice sampling efficiency, \effss, is listed for NS. A total, \eff, and unweighting, \effuw, efficiency are listed
for both algorithms. We report the mean and standard deviation of ten independent runs of the respective
algorithm.}\label{tab:toy_eff}
  \begingroup
\setlength{\tabcolsep}{8pt} 
    \begin{tabular}{lllllll}
      \toprule
Algorithm & \multicolumn{1}{c}{\nlike} & \multicolumn{1}{c}{\effss} & \multicolumn{1}{c}{\nw} &
\multicolumn{1}{c}{\effuw} & \multicolumn{1}{c}{\nequals} & \multicolumn{1}{c}{\eff} \\
      \midrule
\Vegas  & \num{300000} & & \num{300000} & \num{0.004 +- 0.002} & \num{1267 +- 460} & \num{0.004 +- 0.002}\\  
NS  & \num{308755 +- 17505} & \num{0.041 +- 0.003} & \phantom{3}\num{12669 +- 147} & \num{0.273 +- 0.007} & \num{3462 +-
96} & \num{0.0113 +- 0.0009}\\ 
      \bottomrule
    \end{tabular}
  \endgroup
\end{table*}

Tab.~\ref{tab:toy_integral} shows the integral estimates along with the corresponding uncertainty measures. While the
pure Monte Carlo errors are of the same size for both algorithms, there is an additional uncertainty for NS. It carries
an uncertainty on the weights of the sampled points, listed as \deltaw. This arises due to the nature of NS using the
volume enclosed by the live points at each iteration to estimate the volume of the likelihood shell. The variance in
this volume estimate can be sampled, which is reflected as a sample of alternative weights for each dead point in the
sample. Summing up these alternative weight samples gives a spread of predictions for the total integral estimate, and
the standard deviation of these is quoted as \deltaw. This additional uncertainty compounds the familiar statistical
uncertainty, listed as \deltamc for all calculations. In Appendix~\ref{app:uncert}, we present the procedure needed to
combine the two NS uncertainties to quote a total uncertainty, \deltatot, as naively adding in quadrature will
overestimate the true error.
\begin{table}
  \centering
\caption{Comparison of integrals calculated in the toy example with \Vegas and NS, along with the respective
uncertainties.}\label{tab:toy_integral}
  \begingroup
\setlength{\tabcolsep}{8pt} 
    \begin{tabular}{lcccccccc}
      \toprule
Algorithm & I & \deltatot & \deltaw & \deltamc  \\
      \midrule
\Vegas & 1.71 & 0.02 &  & 0.02\\ 
NS & 1.65 & 0.05 & 0.04 & 0.02\\ 
      \bottomrule
    \end{tabular}
  \endgroup
\end{table}

\section{Application to Gluon Scattering}\label{sec:gluon}

As a first application and benchmark for the Nested Sampling algorithm, we consider partonic gluon scattering processes
into three-, four- and five-gluon final states at fixed centre-of-mass energies of $\sqrt{s}=1\,\text{TeV}$. These channels have a complicated phase space structure that is similar to processes with quarks or jets, while the corresponding amplitude expressions are rather straightforward to generate. The fixed initial and final states allow us to focus on the underlying sampling problem. For
regularisation we apply cuts to the invariant masses of all pairs of final state gluons such that
$m_{ij}>\SI{30}{\giga\electronvolt}$ and on the transverse momenta of all final state gluons such that
$p_{\mathrm{T},i}>\SI{30}{\giga\electronvolt}$. The renormalisation scale is fixed to $\mu_R=\sqrt{s}$. The matrix
elements are calculated using a custom interface between \PC and the matrix element generator
\Amegic~\cite{Krauss:2001iv} within the \Sherpa event generator framework~\cite{Sherpa:2019gpd}. Three established
methods are used to provide benchmarks to compare NS to. Principle comparison is drawn to the \HAAG sampler, optimised
for QCD antenna structures~\cite{vanHameren:2002tc}, illustrating the exploration of phase space with the best a priori
knowledge of the underlying physics included. It uses a cut-off parameter of $s_0=\SI{900}{\giga\electronvolt\squared}$.
Alongside this, two algorithms that will input no prior knowledge of the phase space, \emph{i.e.}\ the integrand, are
used; adaptive importance sampling as realised in the \Vegas algorithm~\cite{Lepage:1977sw} and a flat uniform sampler
realised using the \RAMBO algorithm~\cite{Kleiss:1985gy,Platzer:2013esa}. \Vegas remaps the variables of the \RAMBO
parametrisation using 50, 70, 200 bins per dimension for the three-, four-, and five-gluon case, respectively. The grid
is trained in 10 iterations using 100k training points each. Note, the dimensionality of the phase space for $n$-gluon
production is $D=3n-4$, where total four-momentum conservation and on-shell conditions for the external particles are
implicit.

As a first attempt to establish NS in this context, we treat the task of estimating the total and differential cross
sections of the three processes \emph{starting with no prior knowledge} of the underlying phase space distribution. For
the purposes of running \PC we provide the flat \RAMBO sampler as the prior, and the likelihood function provided is the
squared matrix element. In contrast to \HAAG, \PC performs the integration without any decomposition into channels,
removing the need for any multichannel mapping. NS is a flexible procedure, and the objective of the algorithm can be
modified to perform a variety of tasks, a recent example has presented NS for computation of small \pvalues in the
particle physics context~\cite{Fowlie:2021gmr}. To establish NS for the task of phase space integration in this study, a
standard usage of \PC is employed, mostly following default values used commonly in Bayesian inference problems.

The discussion of the application of NS to gluon-scattering processes is split into four parts. Firstly, the
hyperparameters and general setup of \PC are explained in Section~\ref{subsec:hyper}. In Section~\ref{subsec:explore} a
first validation of NS performing the core tasks of (differential) cross-section estimation from weighted events --
against the \HAAG algorithm -- is presented. In Section~\ref{subsec:eff} further information is given to contextualise
the computational efficiency of NS against the alternative established tools for these tasks. Finally a consideration of
unweighted event generation with NS is presented in Section~\ref{subsec:uw}.

\subsection{\PC hyperparameters}\label{subsec:hyper}
\begin{table*}
  \centering
\caption{\PC hyperparameters used for this analysis, parameters not listed follow the \PC defaults.}\label{tab:pc_hyper}
\begin{tabular}{l l l l}
  \toprule
Parameter & \PC name & Value & Description \\
  \midrule
Number of dimensions & \hspace{1cm}\ndim & [5,8,11] & Dimension of sampling space \\
Number of live points & \hspace{1cm}\nlive & 10000 & Resolution of the algorithm \\
Number of repeats & \hspace{1cm}\nrep & $\ndim \times 2$ & Length of Markov chains \\
Number of prior samples & \hspace{1cm}\nprior & \nlive & Number of initial samples from prior \\
Boost posterior & & \nrep & Write out maximum number of posterior samples \\
  \bottomrule
  \end{tabular}
\end{table*}

The hyperparameters chosen to steer \PC are listed in Table~\ref{tab:pc_hyper}. These represent a typical set of choices
for a high resolution run with the objective of producing a large number of posterior samples. The number of live points
is one of the parameters that is most free to tune, being effectively the resolution of the algorithm. Typically \nlive
larger than \ofOrder{1000} gives diminishing returns on accuracy, Bayesian inference usage in particle physics has
previously employed $\nlive=4000$~\cite{Carragher:2021qaj} to provide some context for the choice made in this work. The
particular event generation use case, partitioning the integral into arbitrarily small divisions (differential cross
sections), logically favours a large \nlive (resolution). The number of repeats is a parameter that controls the length
of the slice sampling chains, the value chosen is the recommended default for reliable posterior sampling, whereas
$\nrep=\ndim \times 5$ is recommended for evidence (total integral) estimation. As this study aims to cover both
differential and total cross sections, the smaller value is favoured as there is a strong limit on the overall
efficiency imposed by how many samples are needed to decorrelate the Markov Chains.

An important point to note is in how \PC treats unphysical values of the phase space variables, \emph{e.g.}\ if they
fall outside the fiducial phase space defined by cuts on the particle momenta. This is not an explicit hyperparameter of
\PC, rather how the algorithm treats points with zero likelihood. In both the established approaches and in \PC the
sampling is performed in the unit hypercube, which is then translated to the physical variables which can be evaluated
for consistency and rejected if they are not physically valid. One of the strengths of NS is that the default behavior
is to consider points which return zero likelihood\footnote{~Since \PC operates in log space, to avoid the infinity
associated with $\log(0)$, log-zero is defined as a settable parameter. By default this is chosen to
$-1\times10^{-25}$.} as being excluded at the prior level. During the initial prior sampling phase, unphysical points
are set to log-zero and the sampling proceeds until \nprior initial physical samples have been
obtained. Provided each connected physical region contains some live points after this initial phase, the iterative phase of MCMC sampling will explore up to the unphysical boundary. This effect necessitates a
correction factor to be applied to the integral, derived as the ratio of total initial prior samples to the physically
valid prior samples. In practice the correction factor is found in the \texttt{prior\_info} file written out by \PC. An
uncertainty on this correction can be derived from order statistics~\cite{Fowlie:2020mzs}, however it was found to be
negligibly small for the purposes of this study so is not included. 

Another standout choice of hyperparameter is the chosen value of \nprior. The number of prior samples is an important
hyperparameter that would typically be set to some larger multiple of \nlive in a Bayesian inference context,
$\nprior=10\times\nlive$ would be considered sensible for a broad range of tasks. For the purpose of generating weighted
events, using a larger value would generally be advantageous, however increasing \nprior will strongly decrease the
efficiency in generating \emph{unweighted} events. As the goal is to construct a generator taking an uninformed prior
all the way through to unweighted events, the default value listed is used. However it is notable that this is a
particular feature of starting from an uninformed prior, if more knowledge were to be included in the prior then a
longer phase of prior sampling becomes advantageous. The final parameter noted, the factor by which to boost posterior
samples, has no effect on \PC at runtime. Setting this to be equal to the number of repeats simply writes out the
maximum number of dead points, hence is needed in this scenario. All plots and tables in the remainder of this section
are composed of one single run of \PC with these settings, with the additional entries in Table~\ref{tab:xs}
demonstrating a join of ten such runs.

\subsection{Exploration and Integrals}\label{subsec:explore} 
\begin{figure*}
  \begin{center}
  
  \includegraphics[width=1.0\textwidth]{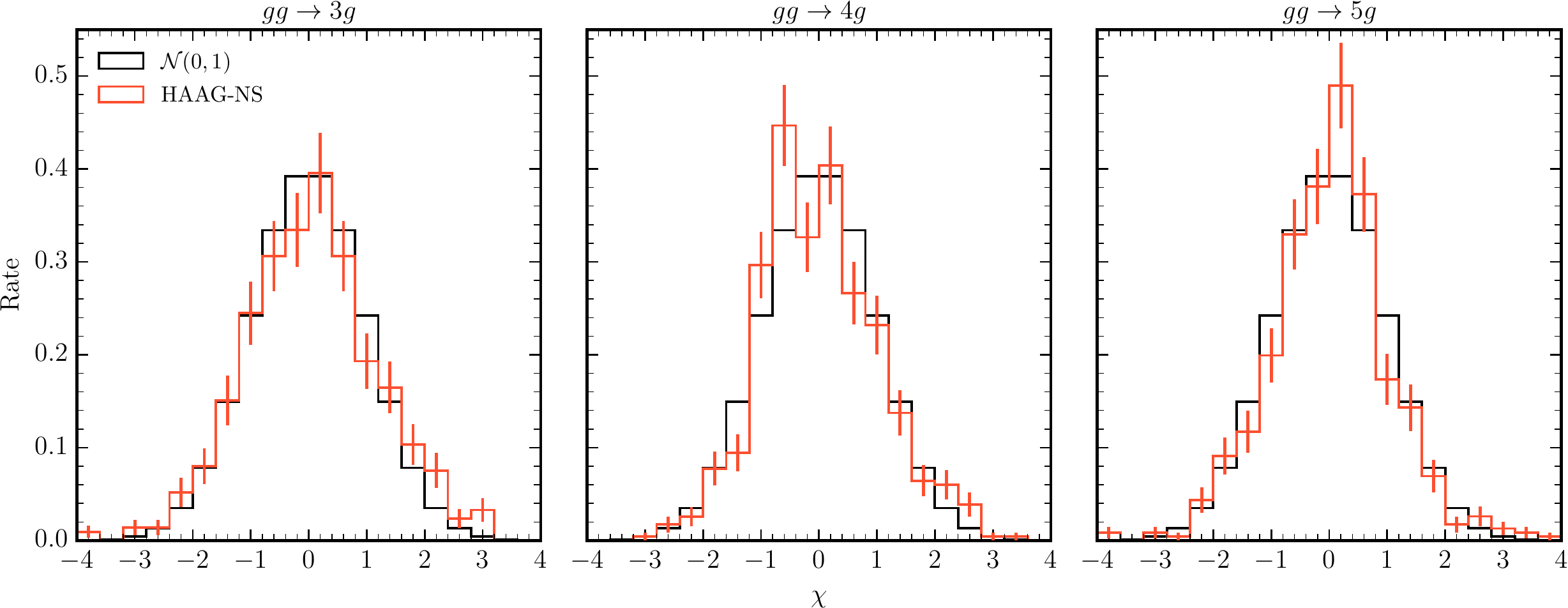}

\caption{Global rate of occurrence of per bin deviation, $\chi$, between \HAAG and NS, for each considered scattering
process. A normally distributed equivalent deviation rate is shown for comparison.}    \label{fig:dev}

\end{center}
\end{figure*}

\begin{figure*}[p]
\begin{center}

\subfloat[Differential cross section binned as a function of the leading jet transverse momentum, for the three
considered processes.\label{fig:pt}]{\includegraphics[width=1.0\textwidth]{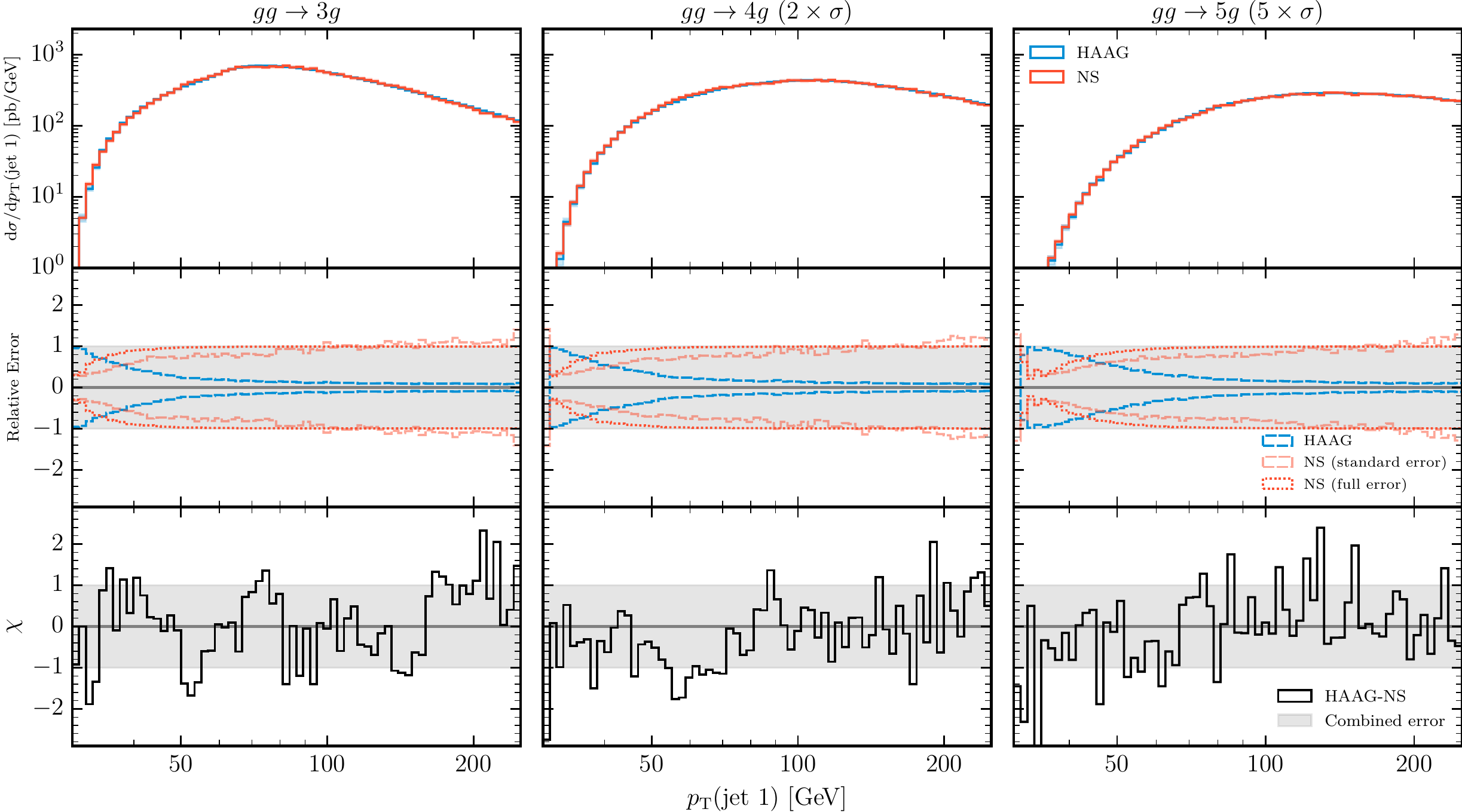}}\\
\subfloat[Differential cross section binned as a function of the separation of the leading two jets in the $(\eta,\phi)$
plane, \emph{i.e.}\ $\Delta R_{12}=\sqrt{\Delta\eta^2_{12} + \Delta\phi^2_{12}}$, for the three considered
processes.\label{fig:dr}]{\includegraphics[width=1.0\textwidth]{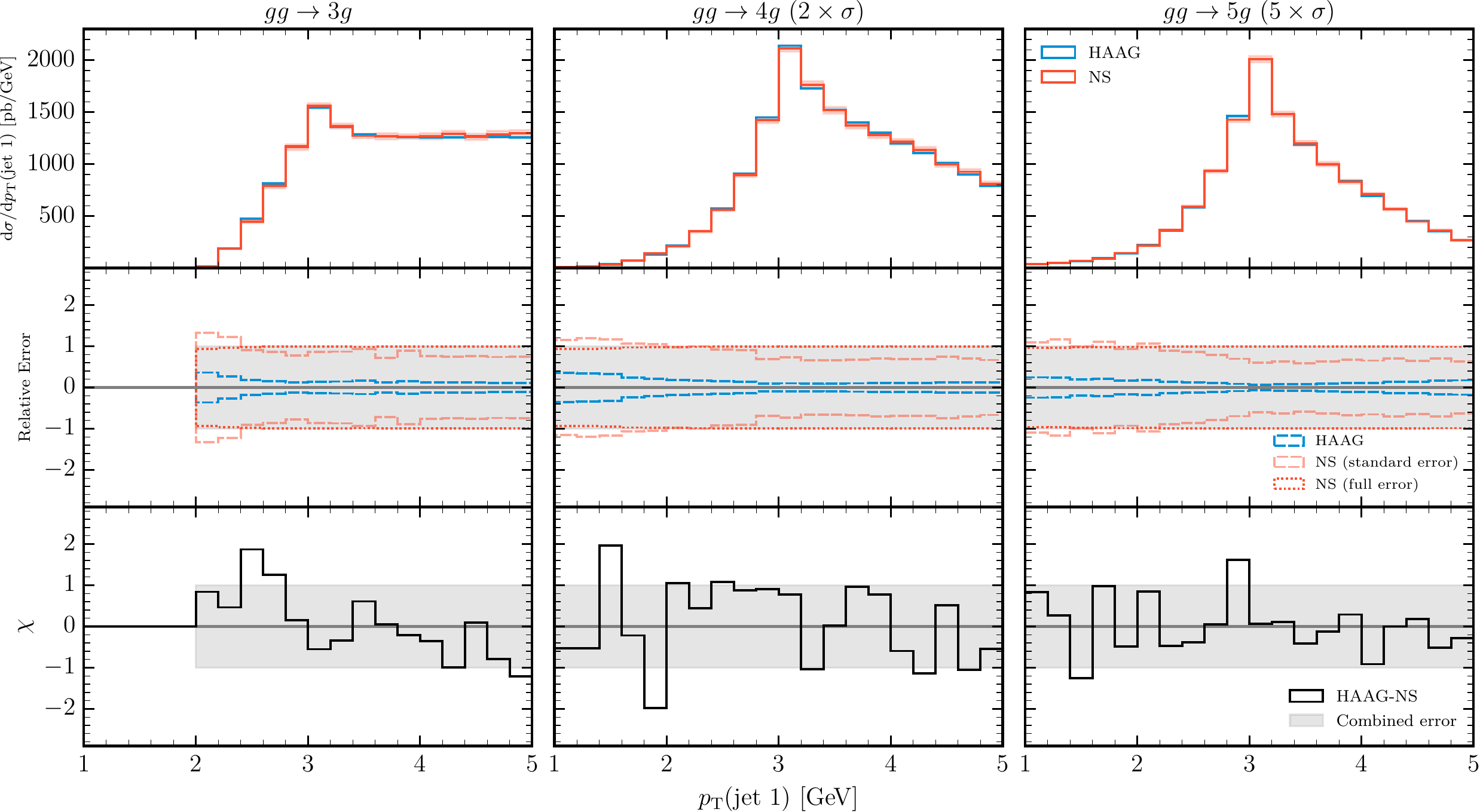}} \caption{Two
example physical differential observables computed with weighted events using the \HAAG and NS algorithms. The top
panels show the physical distributions, the middle panels display the relative component error sources, and the bottom
panel displays the normalised deviation. The deviation plot has been normalised such that $\chi=1$ corresponds to an
expected $1\sigma$ deviation of a Gaussian distribution. Note that for illustrative purposes the cross sections for the
four- and five-gluon processes have been scaled by global factors.}    \label{fig:wobs}

\end{center}

\end{figure*}

Before examining the performance of NS in detail, it is
first important to validate that the technique is capable of fully exploring particle physics phase spaces in these
chosen examples. The key test to validate this is to compare if various differential cross sections calculated with NS
are statistically consistent with the established techniques. To do this, a single NS and \HAAG sample of weighted
events is produced, using approximately similar levels of computational overhead (more detail on this is given in
Section~\ref{subsec:eff}). Both sets of weighted events are analysed using the default MC\_JETS \Rivet
routine~\cite{Bierlich:2019rhm}. \Rivet produces binned differential cross sections as functions of various physical
observables of the outgoing gluons. For each process, the total cross section for the NS sample is normalised to the
\HAAG sample, and a range of fine grained differential cross sections is calculated using both algorithms covering the
following observables; $\eta_{i}$, $y_i$, $p_{\mathrm{T},i}$, $\Delta\phi_{ij}$, $m_{ij}$, $\Delta R_{ij}$,
$\Delta\eta_{ij}$, where $i\neq j$ label the final state jets, reconstructed using the anti-$k_T$
algorithm~\cite{Cacciari:2008gp} with a radius parameter of $R=0.4$ and $p_{\mathrm{T}}>30\,\mathrm{GeV}$. The
normalised difference between the NS and \HAAG differential cross section in each bin can be computed as,
\begin{equation}
  \chi = \frac{d\sigma_{\mathrm{HAAG}} - d\sigma_{\mathrm{NS}}}{ \sqrt{\Delta_{\mathrm{HAAG}}^2 + \Delta_{\mathrm{NS}}^2} } \,,
\end{equation}
in effect this is the differences between the two algorithms normalised by the combined standard deviation. By summing
up this $\chi$ deviation across all the available bins in each process, a test to see if the two algorithms are
convergent within their quoted uncertainties can be performed. Since over 500 bins are populated and considered in each
process, it is expected that the rate of these $\chi$ deviations should be approximately normally distributed. This
indeed appears to hold, and these summed density estimates across all observables are shown in Figure~\ref{fig:dev},
alongside an overlaid normal distribution with mean zero and variance one, ${\cal{N}}(0,1)$, to illustrate the expected
outcome. Two example variables that were used to build this global deviation are also shown; the leading jet \pT in
Fig.~\ref{fig:pt} and $\Delta R_{12}$, the distance of the two leading jets in the $(\eta,\phi)$ plane, in
Fig.~\ref{fig:dr}.

The composition of the quoted uncertainty for the two algorithms differs, demonstrating an important feature of an NS
calculation. For \HAAG, and IS in general, it is conventional to quote the uncertainty as the standard error from the
effective number of fills in a bin. Nested Sampling on the other hand introduces an uncertainty on the weights used to
fill the histograms themselves, effectively giving rise to multiple weight histories that must be sampled to derive the
correct uncertainty on the NS calculation. Details on this calculation are supplied in Appendix~\ref{app:uncert}. In
summary the alternative weight histories give an overlapping measure of the statistical uncertainty, so this effect must
be accounted for in situ alongside taking the standard deviation of the weight histories. To contextualise this, the
middle panels in Fig.~\ref{fig:wobs} show the correct combined uncertainty (using the recipe from
Appendix~\ref{app:uncert}) as a grey band, against the bands derived from the standard error of each individual
algorithm (henceforth \deltamc) as dashed lines, and the complete NS error treatment as a dotted line. The standard
error (dashed) NS band in these panels is a naive estimation of the full NS uncertainty (dotted), however this
illustrates an important point; at the level of fine grained differential observables the NS uncertainty is dominated by
statistics and is hence reducible as one would expect by repeated runs. Based on the example observables we can
initially conclude that whilst both algorithms appear compatible, when using weighted events NS generally has a larger
uncertainty than \HAAG across most of the range (given a roughly equivalent computational overhead). However, further
inspection of the resulting unweighted event samples derived from these weighted samples in the remaining sections
reveals a more competitive picture between the two algorithms.

\begin{table}
  \centering
\caption{Comparison of integrals calculated for the three-, four- and five-gluon processes using \RAMBO, \Vegas, NS and
\HAAG, along with the respective uncertainties.}\label{tab:xs}
  \begingroup
\setlength{\tabcolsep}{8pt} 
\begin{tabular}{l l S[table-format=2.3] c c c c c c c}
      \toprule
Process & Algorithm & \xs & \deltatot & \deltaw & \deltamc  \\
      \midrule
$3-$jet & \RAMBO & 24.580 & 0.191 &  & 0.191\\ 
& \Vegas & 24.807 & 0.017 &  & 0.017\\ 
& NS & 24.669 & 0.467 &  0.484 & 0.100\\ 
& NS ${(\times 10)}$ & 24.888 & 0.145 & 0.150 & 0.030\\ 
& \HAAG & 24.840 & 0.017 &  & 0.017\\ 
      \midrule
$4-$jet & \RAMBO & 9.876 & 0.107 &  & 0.107\\ 
& \Vegas & 9.849 & 0.009 &  & 0.009\\ 
& NS  & 9.837 & 0.194 & 0.196 & 0.036\\ 
& NS ${(\times 10)}$ & 9.778 & 0.064 & 0.066 & 0.011\\ 
& \HAAG & 9.853 & 0.006 &  & 0.006\\ 
      \midrule
$5-$jet & \RAMBO & 2.644 & 0.024 &  & 0.024\\ 
& \Vegas & 2.680 & 0.003 &  & 0.003\\ 
& NS & 2.612 & 0.051 & 0.048 & 0.009\\ 
& NS ${(\times 10)}$& 2.667 & 0.017 & 0.017 & 0.003\\ 
& \HAAG & 2.685 & 0.001 &  & 0.001\\ 
      \bottomrule
    \end{tabular}
  \endgroup
\end{table}
The estimates of the total cross sections, derived from the sum of weighted samples, provided in Table~\ref{tab:xs},
give an alternative validation that NS is sufficiently exploring the phase space by ensuring that compatible
estimates of the cross sections are produced between all the methods reviewed in this study. The central estimates of
the total cross sections are generally consistent within the established error sources for all calculations considered.
In this table the components of the error calculation for NS are listed separately; \deltaw being the standard deviation
resulting from the alternative weight histories and \deltamc being the standard error naively taken from the mean of the
alternative NS weights. In contrast to the differential observables, the naive counting uncertainty is small so has
negligible effect at the level of total cross sections. In summary, for a total cross section the spread of alternative
weight histories gives a rough estimate of the total error, whereas for a fine grained differential cross section the
standard error dominates. The way to correctly account for the effect of counting statistics within the weight histories
is given in Appendix~\ref{app:uncert}. 

Repeated runs of NS will reduce these uncertainties. The \anesthetic package~\cite{Handley:2019mfs} is used to analyse
the NS runs throughout this paper, and contains a utility to join samples. Once samples are joined consistently into a
larger sample, the uncertainties can be derived as already detailed. The result of joining 10 equivalent NS runs with
the previously motivated hyperparameters is also listed in Table~\ref{tab:xs}. Joining 10 runs affects the \deltatot for
NS in two ways; reducing the spread of weighted sums composing \deltaw (\emph{i.e.}\ reducing \deltamc), and reducing
the variance of distribution for each weight itself (\emph{i.e.}\ the part of \deltaw that does not overlap with
\deltamc). The former is reduced by simply having an increased size of samples produced, increasing the number of
effective fills by a factor of $\sim$10 in this case, with the latter reduced due to the increased effective number of
live points used for the volume estimation. 

\subsection{Efficiency of Event Generation}\label{subsec:eff}

An example particle physics workflow on this gluon scattering problem would be to take \HAAG as an initial mapping of
the phase space (effectively representing the best prior knowledge of the problem), and using \Vegas to refine the
proposal distribution to optimally efficiently generate weighted events. Of the three existing tools presented in this
study for comparison (\HAAG, \RAMBO, and \Vegas), NS bears most similarity to \Vegas, in that both algorithms learn the
structure of the target integrand. To this end an atypical usage of \Vegas is employed, testing how well \Vegas could
learn a proposal distribution from an uninformed starting point (\RAMBO). This is equivalent to how NS was employed,
starting from an uninformed prior (\RAMBO) and generating posterior samples via Nested Sampling. It was motivated so far
that roughly similar computational cost was used for the previous convergence checks, and that the hyperparameters of
\PC were chosen to emphasise efficient generation of unweighted events. In what follows, we analyse more precisely this
key issue of computational efficiency.

The statistics from a single run of the four algorithms for the three selected processes is listed in
Table~\ref{tab:eff}. NS is non deterministic in terms of number of matrix element evaluations (\nlike), instead
terminating from a pre determined convergence criterion of the integral. \HAAG, \Vegas, \RAMBO are all used to generate
exactly 10M weighted events. The chosen \PC hyperparameters roughly align the NS method with the other three in terms of
computational cost. One striking difference comes from the Markov Chain nature of NS. Default usage only retains a
fraction of the total \like evaluations, inversely proportional to \nrep. This results in a smaller number of retained
weighted events, \nw, than the number of \like evaluations, \nlike, for NS. However the retained weighted events by
construction match the underlying distribution much closer than the other methods, resulting in a higher unweighting
efficiency, \effuw, for the NS sample. Exact equal-weight unweighting can be achieved by accepting events with a
probability proportional to the share of the sample weight they carry, this operation is performed for all samples of
weighted events and the number of retained events is quoted as \nequals. NS as an unweighted event generator has some
additional complexity due to the uncertainty in the weights themselves, this is given more attention in
Section~\ref{subsec:uw}.

Due to differences in \nlike between NS and the other methods, it is most effective to compare the total efficiency in
producing unweighted events, $\eff=\nequals/\nlike$. \RAMBO as the baseline illustrates the performance one would
expect, inputting no prior knowledge and not adapting to any acquired knowledge. As such \RAMBO yields a tiny \eff.
\HAAG represents the performance using the best state of prior knowledge but without any adaptation, in these tests this
represents the best attainable \eff. \Vegas and NS start from a similar point, both using \RAMBO as an uninformed state
of prior knowledge, but adapting to better approximate the phase space distribution as information is acquired. \Vegas
starts with a higher efficiency than NS for the $3-$gluon process, but the \Vegas efficiency drops by approximately an
order of magnitude as the dimensionality of phase space is increased to the $5-$gluon process. NS maintains a consistent
efficiency of approximately a percent, competitive with the consistent approximately three percent efficiency obtained
by \HAAG.

\begin{table*}
  \centering
\caption{Comparison of the four algorithms for the three processes in terms of size of event samples produces. \nlike
gives the number of matrix element evaluations, \nw the number of weighted events, $N_{W,\mathrm{eff}}$ the effective
number of weighted events and \nequals the derived number of equal-weight events. A MC slice sampling efficiency,
\effss, is listed for NS. A total, \eff, and an unweighting, \effuw, efficiency are listed for all
algorithms.}\label{tab:eff}
  \begingroup
\setlength{\tabcolsep}{8pt} 
\begin{tabular}{l l S[table-format=2.2] c S[table-format=2.2] S[table-format=1.5] S[table-format=1.4]
S[table-format=1.5]}
      \toprule
Process & Algorithm & \nlike $^{(\times10^6)}$ &\effss & \nw $^{(\times10^6)}$ & \effuw & \nequals $^{(\times10^6)}$ &
\eff \\
      \midrule
$3$-jet & \RAMBO  & 10.00 &   & 10.00   & 0.0001 & 0.001 & 0.0001\\ 
& \Vegas  & 10.00 &   & 10.00 &  0.02 & 0.20 & 0.02\\  
& NS  & 6.43 & 0.03 & 0.17  & 0.37 & 0.06 & 0.01\\ 
& \HAAG  & 10.00 &   & 10.00  & 0.03 & 0.29 & 0.03\\
      \midrule 
$4$-jet & \RAMBO  & 10.00 &  &  10.00 &  0.00003 & 0.0003 & 0.00003\\ 
& \Vegas  & 10.00 &   & 10.00  &  0.005 & 0.049 & 0.005\\ 
& NS  & 7.94 & 0.02 & 0.19  & 0.43 & 0.08 & 0.01\\ 
& \HAAG  & 10.00 &   & 10.00  & 0.02 & 0.23 & 0.02\\
      \midrule 
$5$-jet & \RAMBO  & 10.00 &  &  10.00 &  0.00004 & 0.0004 & 0.00004\\ 
& \Vegas  & 10.00 &   & 10.00 &  0.001 & 0.013 & 0.001\\ 
& NS  & 9.17 & 0.02 & 0.19 &  0.44 & 0.08 & 0.01\\ 
& \HAAG  & 10.00 &   & 10.00 &  0.03 & 0.25 & 0.03 \\
      \bottomrule
    \end{tabular}
  \endgroup
\end{table*}

\begin{figure}
  \includegraphics[width=0.4\textwidth]{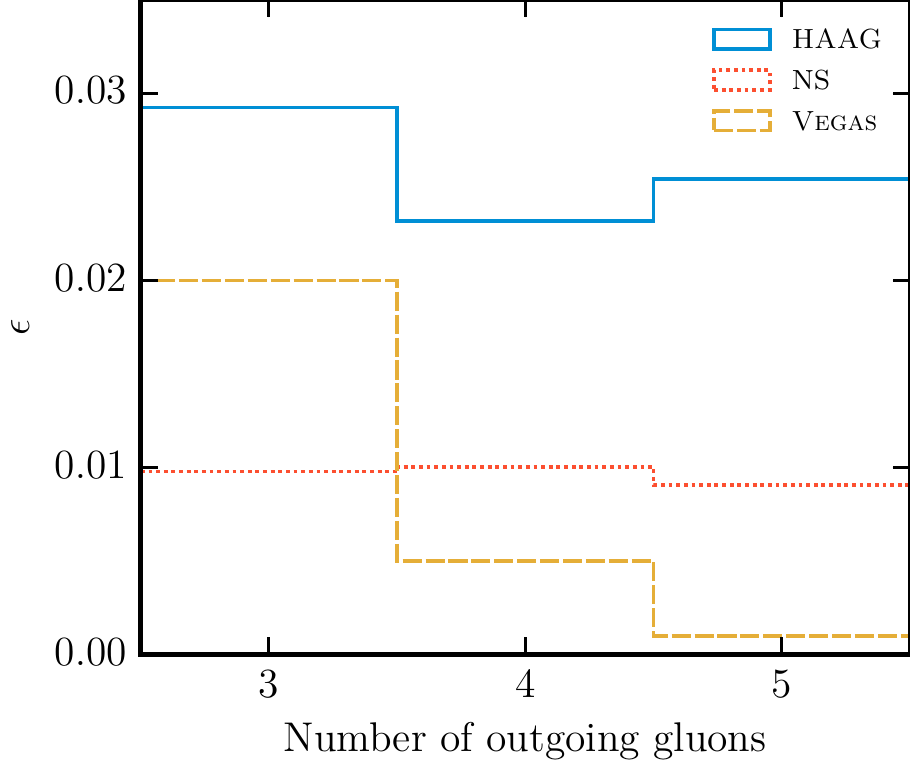}
\caption{Visualisation of the efficiencies listed in Table~\ref{tab:eff}.}    \label{fig:eff}
\end{figure}

As the key point of comparison for this issue is the efficiency, \eff, this is highlighted with an additional
visualisation in Fig.~\ref{fig:eff}. The scaling behavior of the efficiency of each algorithm as a function of the
number of outgoing gluons (corresponding to an increase in phase space dimensionality) is plotted for NS, \HAAG and
\Vegas. From the same starting point, NS and \Vegas can both learn a representation of the phase space, and do so in a
way that yields a comparable efficiency to the static best available prior knowledge in \HAAG. As the dimensionality of
the space increases it appears that \Vegas starts to suffer in how accurately it can learn the mapping, however NS is
still able to learn the mapping in a consistently efficient manner.   

\subsection{Unweighted Event Generation}\label{subsec:uw} The fact that NS leads to a set of alternative weight
histories poses a technical challenge in operating as a generator of unweighted events in the expected manner. Exact
unweighting, compressing the weighted sample to strictly equally weighted events leads to a different set of events
being accepted for each weight history. Representative yields of unweighted events can be calculated as shown in
Table~\ref{tab:eff} using the mean weight for each event, but the resulting differential distributions will
underestimate the uncertainty if this is quoted simply as the standard error in the bin, as described in
Appendix~\ref{app:uncert}. The correct uncertainty recipe can be propagated through naively, by separately unweighting
each weight history, however this requires saving as many event samples as required weight variations. Partial
unweighting is commonly used in HEP event generation to allow a slight deviation from strict unit weights, to increase
efficiency in practical settings. A modification to the partial unweighting procedure could be used to propagate the
spread of weights to variations around accepted, approximate unit weight, events. 

\begin{figure*}
  \includegraphics[width=1.0\textwidth]{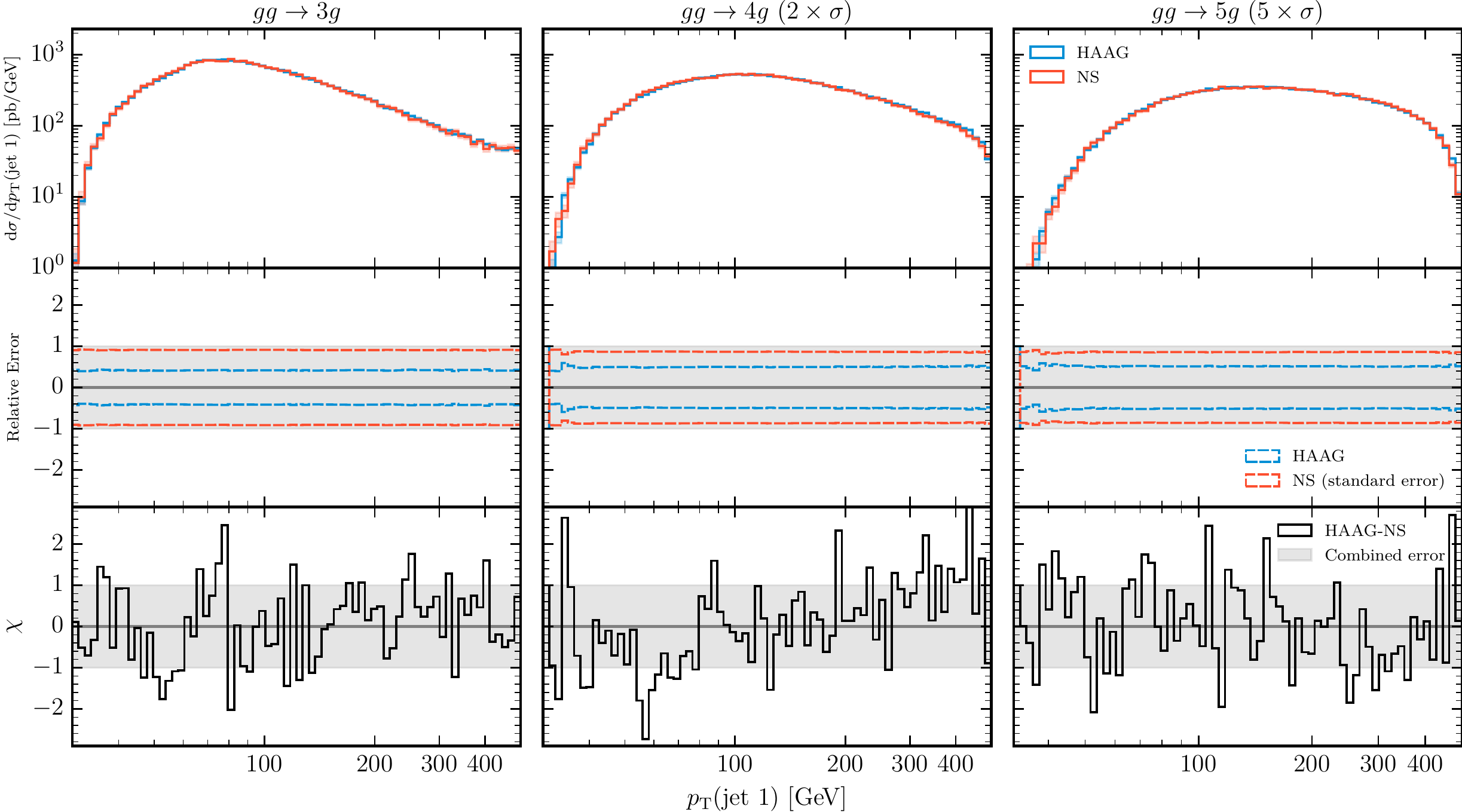}
\caption{The equivalent leading jet transverse momentum observable as calculated in Fig.~\ref{fig:pt}, using an exact
unit weight compression of the same samples. A modified version of the default MC\_JETS routine has been used to extend
the \pT range shown.}    \label{fig:ptuw}
\end{figure*}

To conclude the exploration of the properties of NS as a generator for particle physics, a representative physical
distribution calculated from a sample of exact unit-weight events is shown in Figure~\ref{fig:ptuw}. This sample is
derived from the same weighted sample described in Table~\ref{tab:eff} and previously presented as a weighted event
sample in Figure~\ref{fig:pt}. The full set of NS variation weights is used to calculate the mean weight for each event,
which is used to unweight the sample, for the chosen observable this is a very reasonable approximation as the fine
binning means the standard error is the dominant uncertainty. The range of the leading jet transverse momenta has been
extended into the tail of this distribution by modifying the default \Rivet routine. This distribution largely reflects
the information about the total efficiency previously illustrated in Figure~\ref{fig:eff}, projected onto a familiar
differential observable. The total efficiency, \eff, was noted as being approximately one percent from NS, compared to
approximately three percent from \HAAG across all processes. If the total number of matrix element evaluations, \nlike,
were to be made equal across all algorithms and processes, the performance would be further consistent. 

\section{Future research directions}\label{sec:directions} Throughout Sec.~\ref{sec:gluon}, the performance of Nested
Sampling in the context of particle physics phase space sampling and event generation was presented. A single choice of
hyperparameters was made, effectively performing a single NS run as an entire end-to-end event generator; starting from
zero knowledge of the phase space all the way through to generating unweighted events. Simplifying the potential options
of NS to a single version of the algorithm was a deliberate choice to more clearly illustrate the performance of NS in
this new context, using the same settings for multiple tasks gives multiple orthogonal views on how the algorithm
performs. However this was a limiting choice, NS has a number of variants and applications that could more effectively
be tuned to a subset of the tasks presented. Some of the possible simple alterations -- such as increasing \nprior to
improve weighted event generation at the expense of unweighting efficiency -- were motivated already in this paper. In
this section we outline four broad topics that extend the workflow presented here, bringing together further ideas from
the worlds of Nested Sampling and phase space exploration.

\subsection{Physics challenges in event generation}
The physical processes studied in this work, up to $5-$gluon scattering problems, are representative of the complexity
of phase space calculation needed for the current precision demands of the LHC experiment
collaborations~\cite{ATLAS:2021yza}. However part of the motivation for this work, and indeed the broader increased
interest in phase space integration methods, is due to the impending breaking point current pipelines face under the
increased precision challenges of the HL-LHC programme. Firstly we observe that the phase space dimensionality of the
highest multiplicity process studied here is 11. In broader Bayesian inference terms this is rather small, with NS being
typically used for problems \ofOrder{10} to \ofOrder{100} dimensions, where it is uniquely able to perform numerical
integration without approximation or strictly matching prior knowledge. The \PC implementation is styled as
\emph{next-generation} Nested Sampling, designed to have polynomial scaling with dimensionality aiming for robust
performance as inference is extended to \ofOrder{100} dimensions. Earlier implementations of NS, such as
\MN~\cite{Feroz:2008xx}, whilst having worse dimensional scaling properties, may be a useful avenue of investigation for
the lower dimensional problems considered in this paper.

This work validated NS in a context where current tools still can perform the required tasks, albeit at times at immense
computational costs. Requirements from the HL-LHC strain the existing LHC event generation pipeline in many ways and
pushing the sampling problem to higher dimensions is no exception~\cite{Campbell:2022qmc}. Importance Sampling becomes
exponentially more sensitive to how close the proposal distribution matches the target in higher dimensions, a clear
challenge for particle physics in two directions; multileg processes rapidly increasing the sampling
dimension~\cite{Hoche:2019flt} and corresponding radiative corrections (real and virtual) make it increasingly hard to
provide an accurate proposal, \emph{e.g.}\ through the sheer number of phase space channels
needed and by having to probe deep into singular phase space regions~\cite{Gleisberg:2007md}. We propose that NS is an excellent complement to further investigation on both these
fronts. The robust dimensional scaling of NS illustrated against \Vegas in Figure~\ref{fig:eff} encapsulates both solid
performance with increasing dimension, and the adherence to an uninformed prior whilst still attaining this scaling is
promising for scenarios where accurate proposals are harder to construct. 

\subsection{Using prior knowledge}
Perhaps the most obvious choice that makes the application here stylised is in always starting from an uninformed prior
state of knowledge. Using Equations~\eqref{eq:ps_samp} and~\eqref{eq:evidence}, the cross section integral with a phase
space mapping was motivated as being exactly the Bayesian evidence integral with a choice of prior. To that end there is
no real distinction between taking the non-uniform \HAAG distribution as the prior instead of the flat \RAMBO density
that was used in this study. In this respect NS could be styled as learning an additional compression to the posterior
distribution, refining the static proposal distributions typically employed to initiate the generation of a phase space
mapping (noting that this is precisely what \Vegas aims to do in this context).

Naively applying a non-flat mapping exposes the conflicting aims at play in this set of problems however; efficiently
generating events from a strongly peaked distribution, and generating high statistics estimates of the tails of the same
distribution. Taking a flat \RAMBO prior is well suited to the latter problem, whereas taking a \HAAG prior is better
suited to the former. One particular hyperparameter of \PC that was introduced can be tuned to this purpose; the number
of prior samples, \nprior. If future work is to use a non flat, partially informed starting point, increasing \nprior
well above the minimum (equal to the number of live points required) used in this study would be needed. A more complete
direction for further work would be to investigate the possibility of mixing multiple proposal
distributions~\cite{supernest,supernestproj}.

As a demonstration, we again apply NS to the toy example of Sec.~\ref{sec:toy_example} but this time using a non-uniform
prior distribution. While a good prior would be an approximation of the target distribution, we choose to purposely miss
an important feature of the target, the straight line segment, that the sampler still has to explore. Considering that
in HEP applications the prior knowledge may be encoded in the mixture distributions of a multi-channel importance
sampler, this is an extreme version of a realistic situation. As typically the number of channels grows dramatically
with increasing final-state particle multiplicity, \emph{e.g.}\ factorially when channels correspond to the topologies
of contributing Feynman diagrams, one might choose to disable some sub-dominant channels in order to avoid a
prohibitively large set of channels. However, this would lead to a mis-modelling of the target in certain phase-space
regions.

\begin{figure}
    \begin{center}
    
    \includegraphics[width=0.5\textwidth]{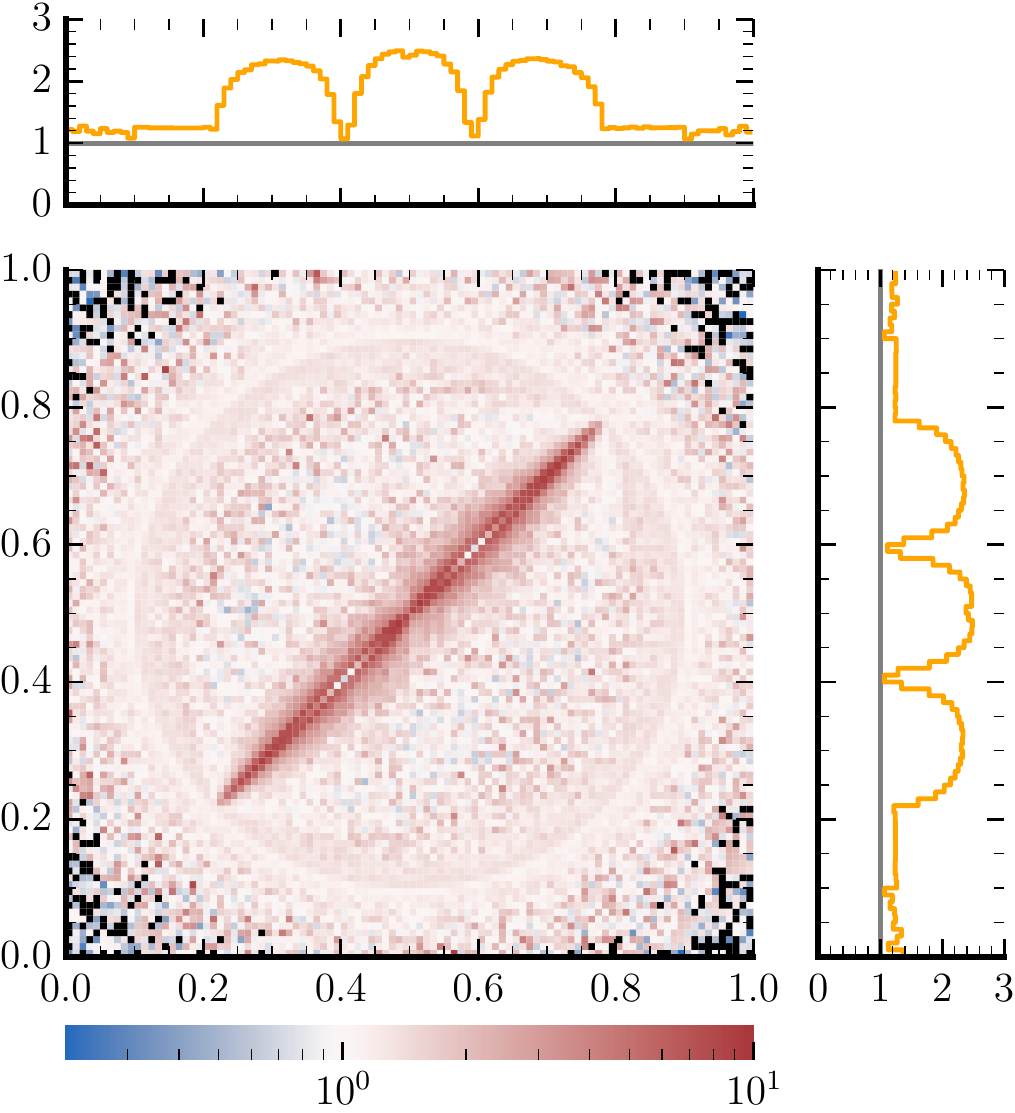}
\caption{The ratio of the target function of the two-dimensional toy example and the probability density function of \PC
using a non-uniform prior distribution. 
Black histogram bins have not been filled by any data due to limited sample size.}
    \label{fig:toy_prior}
  \end{center}
\end{figure}

Here we use only the ring part of the target, truncated on a circle that covers the unit hypercube, as our prior.
Without an additional coordinate transformation this prior would not be of much use for \Vegas as the line part remains
on the diagonal. To sample from the prior, we first transform to polar coordinates. Then we sample the angle uniformly
and the radial coordinate using a Cauchy distribution truncated to the interval $(0, 1/\sqrt{2}]$. In order to have good
coverage of the tails, despite the strongly peaked prior, we increase \nprior to $50\times n_\text{live}$. This results
in a total efficiency of $\eff = \num{0.037+- 0.004}$, more than three times the value obtained with a uniform prior,
\emph{cf.}\ Tab.~\ref{tab:toy_eff}. While the unweighting efficiency reduces to $\effuw = \num{0.17 +- 0.02}$, the slice
sampling efficiency increases to $\effss = \num{0.216 +- 0.007}$. In Fig.~\ref{fig:toy_prior} we show the ratio between
the target function and the \PC sampling distribution. Compared to Fig.~\ref{fig:toy_polychord}, the ratio has a smaller
range of values. Along the peak of the ring part of the target function, the ratio is approximately one. The largest
values can be found around the line segment with \PC generating up to ten times less samples than required by the target
distribution. It can be concluded that even with an intentionally poor prior distribution, \PC benefits from the prior
knowledge in terms of efficiency and still correctly samples the target distribution including the features absent from
the prior.

\subsection{Dynamic Nested Sampling}
In addition to using a more informed prior to initiate the Nested Sampling process, a previous NS run can be used to
further tune the algorithm itself to a particular problem. This is an existing idea in the literature known as dynamic
Nested Sampling~\cite{dyn_ns}. Dynamic NS uses information acquired about the likelihood shells in a previous NS run to
varying the number of live points \emph{dynamically} throughout the run. This results in a more efficient allocation of
the computation towards the core aim of compressing the prior to the posterior. We expect that this would only increase
the efficiency of the unweighting process, as the density of weighted events would be trimmed to even more closely match
the underlying phase space density. Dynamic Nested Sampling naturally combines with the proposal of \emph{using prior
knowledge} to make a more familiar generator chain, however one that is driven primarily by NS. This mirrors the current
established usage of \Vegas in this context; using \Vegas to refine the initial mapping by a redistribution of the input
variables, to more efficiently generate from the acquired mapping.

\subsection{Connection to modern Machine Learning techniques}
There has been a great deal of recent activity coincident to this work, approaching similar sets of problems in particle
physics event generation using modern Machine Learning (ML) techniques~\cite{Butter:2022rso}. Much of this work is still
exploratory in nature, and covers such a broad range of activity that comprehensively reviewing the potential for
combining ML and NS is beyond the scope of this work. It is however clear that there is strong potential to include NS
into a pipeline that modern ML is already aiming to optimise. To that aim, we identify a particular technique that has
been studied previously in the particle physics context; using Normalising Flows to train phase space
mappings~\cite{Bothmann:2020ywa,Gao:2020vdv,Gao:2020zvv}. In spirit a flow based approach, training an invertible
probabilistic mapping between prior and posterior, bears a great deal of similarity to the core compression idea behind
Nested Sampling. The potential in dovetailing Nested Sampling with a flow based approach has been noted in the NS
literature~\cite{Alsing:2021wef}, further motivating the potential for synergy here.

The ability of NS to construct mappings of high dimensional phase spaces without needing any strong prior knowledge, can
be motivated as being an ideal forward model with which to train a Normalising Flow. In effect this replaces the
generator part of the process with an importance sampler, whilst still using NS to generate the mappings. This is
particularly ideal in this context, as the computational overhead required to decorrelate the Markov Chains imposes a
harsh limit on the efficiency of a pure NS based approach. Combining these techniques in this way could retain the
desirable features of both and serve to mitigate the ever increasing computational demands of energy frontier particle
physics.

We close by noting that also in the area of lattice field theory Normalising Flows have recently attracted attention,
see \emph{e.g.}\ \cite{DelDebbio:2021qwf,Hackett:2021idh}, to address the sampling of multimodal target function. We
envisage that also in these applications Nested Sampling could be applied. 

\section{Conclusions}\label{sec:conc}

The establishing study presented here had two main aims. Firstly to introduce the technique of Nested Sampling, applied
to a realistic problem, to researchers in the particle physics community. Secondly to provide a translation back to
researchers working on Bayesian inference techniques, presenting an important and active set of problems in particle
physics that Nested Sampling could provide a valuable contribution to. The physical example presented used \PC to
perform an end-to-end generation of events without any input prior knowledge. This is a stylised version of the event
generator problem, intended to validate Nested Sampling in this new context and demonstrate some key features. For the
considered multi-gluon production processes Nested Sampling was able to learn a mapping in an efficient manner that
exhibits promising scaling properties with phase space dimension. We have outlined some potential future research
directions; highlighting where the strengths of this approach could be most effective, and how to embed Nested Sampling
in a more complete event generator workflow. Along these lines, we envisage an implementation of the Nested Sampling
technique for the \Sherpa event generator framework~\cite{Sherpa:2019gpd}, possibly also supporting operation on
GPUs~\cite{Bothmann:2021nch}. This will provide additional means to address the computing challenges for event
generation posed by the upcoming LHC runs. 

\section*{Acknowledgments}
This work has received funding from the European Union's Horizon 2020 research and innovation programme as part of the
Marie Sk\l{}odowska-Curie Innovative Training Network MCnetITN3 (grant agreement no. 722104). SS and TJ acknowledge
support from BMBF (contract 05H21MGCAB). SS acknowledges funding by the Deutsche Forschungsgemeinschaft (DFG, German
Research Foundation) - project number 456104544. WH and DY are funded by the Royal Society.

This work was performed using resources provided by the Cambridge Service for Data Driven Discovery (CSD3) operated by
the University of Cambridge Research Computing Service (\url{www.csd3.cam.ac.uk}), provided by Dell EMC and Intel using
Tier-2 funding from the Engineering and Physical Sciences Research Council (capital grant EP/T022159/1), and DiRAC
funding from the Science and Technology Facilities Council (\url{www.dirac.ac.uk}).

\appendix
\section{Uncertainties in Nested Sampling}\label{app:uncert} Typical Nested Sampling literature focuses on two main
sources of uncertainty; an uncertainty on the weights of the dead points due to the uncertainty in the volume
contraction at each iteration, and an uncertainty on the overall volume arising from the path the Markov Chain takes
through the space to perform each iteration. The former source is what we consider in this work, and can be calculated
as a sample of weights for each dead point using \anesthetic. The latter source can be estimated using the \nestcheck
package~\cite{Higson:2018cqj}, the method presented here uses combinations of multiple runs to form integral estimates
meaning the best strategy to minimise this effect is already baked in. Further use cases would benefit from more
thorough cross checks using \nestcheck.
\par
The usual source of uncertainty in a binned histogram in particle physics comes from the standard error. Importance
Sampling draws sample events with associated weights $w_i$, with the sum of these sample weights giving the estimated
cross section in a bin. The effective number of fills in a bin using weighted samples is,
\begin{equation}\label{eq:neff}
  N= \frac{\big( \sum_i w_{i} \big)^2}{\sum_i w_{i}^2}\,.
\end{equation}
The inverse square root of $N$ then constitutes the standard error on the cross section in the bin. In practice this
means that the standard deviation of an integral estimated with Importance Sampling can be quoted as $\deltamc =
\sqrt{\sum_i (w_{i}^2)} $. In typical NS applications this is significantly smaller than the previously mentioned
sources, and thus often not considered. However, when using NS as a phase space event generator for finely binned
differential observables, the statistical uncertainty can become a significant effect so must be taken into account.
Adding the standard error to the weight uncertainty in quadrature is a suitable upper bound for the NS uncertainty but
is found to overestimate the uncertainty in some bins. While the standard error gives a measure of the spread of weights
around the mean weight in a bin, alternative weights from the sampling history in NS also give an overlapping measure of
this.
\par
To correctly account for the statistical error in this context a revised recipe is needed. The following proposed
procedure reweights the alternative weight samples to account for the spread of the resulting effective fills in each
bin. The effective number of entries in a bin arising from a NS run can be written as,
\begin{equation}\label{eq:multi}
  N_j= \frac{\big( \sum_i w_{j,i} \big)^2}{\sum_i w_{j,i}^2}\,,
\end{equation}
where $i$ indexes the number of weighted samples in each bin, and $j$ indexes the alternative weights. The result of the
$j$ sampled weight variations is a set of $j$ different effective counts in each bin. These counts can be modelled as
$j$ trials of a multinomial distribution with $j$ categories, written as,
\begin{equation}
  \Pg{N}{\alpha}=\frac{j!}{\prod_j N_j!}\prod_j\alpha^{N_j}_j\,,
\end{equation}
where a probability of sampling each category, $\alpha_j$, has been introduced. The desired unknown distribution of
$\alpha_j$ can be found using Bayes theorem to invert the arguments. If an uninformative conjugate prior to the
multinomial distribution is used, the Dirichlet distribution, the desired inverted probability can also be written in
the form of a Dirichlet distribution,
\begin{equation}
  \Pg{\alpha}{N}= \Gamma\bigg(\sum_j N_j\bigg) \prod_j \frac{\alpha_j^{N_j-1}}{\Gamma(N_j)}\,.
\end{equation}
A sample vector of $\alpha_j$ from this Dirichlet distribution, will give a probability of observing each category
$N_j$. This probability can be used to weight the categories giving a weighted set of effective number of fills,
$\{\alpha_j N_j\}$. This considers each alternative weight sample as a discrete sample from an underlying continuous
distribution $N_j$ is sampled from. The set of weighted effective fills can be used to quote a weighted set of samples
of the bin cross section by multiplying by the square of the sum of the weights, $\{\sigma_j\}=\{\alpha_j N_j
\sum_i(w_{j,i}^2)\}$. The estimated cross section in the bin is then the expected value of this set, $\sigma =
E[\sigma_j]$, and the total standard deviation on this cross section is derived from the variance, $\deltatot=
(\mathrm{Var}[\sigma_j])^2$.

\bibliographystyle{JHEP}
\bibliography{references}

\end{document}